\documentclass[sts]{imsart}

\RequirePackage{amsthm,amsmath,amsfonts,amssymb}
\RequirePackage[authoryear]{natbib}
\RequirePackage[colorlinks,citecolor=blue,urlcolor=blue]{hyperref}
\RequirePackage{graphicx}
\RequirePackage{soul,cancel} 
\usepackage{pdfpages}

\startlocaldefs

\def\be{\boldsymbol{e}}

\def\br{\boldsymbol{r}}

\def\bu{\boldsymbol{u}}
\def\bv{\boldsymbol{v}}

\def\bx{\boldsymbol{x}}
\def\by{\boldsymbol{y}}
\def\bz{\boldsymbol{z}}
\def\bA{\boldsymbol{A}}

\def\bC{\boldsymbol{C}}
\def\bD{\boldsymbol{D}}

\def\bG{\boldsymbol{G}}
\def\bH{\boldsymbol{H}}
\def\bI{\boldsymbol{I}}

\def\bL{\boldsymbol{L}}
\def\bM{\boldsymbol{M}}

\def\bO{\boldsymbol{O}}
\def\bP{\boldsymbol{P}}

\def\bR{\boldsymbol{R}}
\def\bS{\boldsymbol{S}}
\def\bT{\boldsymbol{T}}

\def\bV{\boldsymbol{V}}
\def\bW{\boldsymbol{W}}
\def\bX{\boldsymbol{X}}

\def\bZ{\boldsymbol{Z}}

\def\bbeta{\boldsymbol{\beta}}

\def\bvarepsilon{\boldsymbol{\varepsilon}}

\def\bdeta{\boldsymbol{\eta}}
\def\btheta{\boldsymbol{\theta}}

\def\bmu{\boldsymbol{\mu}}

\def\bxi{\boldsymbol{\xi}}

\def\bSigma{\boldsymbol{\Sigma}}

\def\bOmega{\boldsymbol{\Omega}}
\def\tr{\mbox{tr}}
\def\diag{\mbox{diag}}

\def\bzero{\boldsymbol{0}}

\def\sigsqH{\sigma^2}

\endlocaldefs

\begin{document}

\begin{frontmatter}
\title{Restricted Maximum Likelihood Estimation in Generalized Linear Mixed Models}
\runtitle{Restricted Maximum Likelihood Estimation in Generalized Linear Mixed Models}

\begin{aug}
\author[A]{\fnms{Luca Maestrini,}~\snm{}\ead[label=e1]{luca.maestrini@anu.edu.au}}
\author[B]{\fnms{Francis K.C. Hui}~\snm{}\ead[label=e2]{francis.hui@anu.edu.au}}
\and
\author[C]{\fnms{Alan H. Welsh}~\snm{}\ead[label=e3]{alan.welsh@anu.edu.au}}
\address[A]{Luca Maestrini is Lecturer, Research School of Finance, Actuarial Studies and Statistics, The Australian National University, Canberra, Australia\printead[presep={\ }]{e1}.}
\address[B]{Francis K.C. Hui is Associate Professor, Research School of Finance, Actuarial Studies and Statistics, The Australian National University, Canberra, Australia\printead[presep={\ }]{e2}.}
\address[C]{Alan H. Welsh is E.J. Hannan Professor of Statistics, Research School of Finance, Actuarial Studies and Statistics, The Australian National University, Canberra, Australia\printead[presep={\ }]{e3}.}
\end{aug}

\begin{abstract}
Restricted maximum likelihood (REML) estimation is a widely accepted and frequently used method for fitting linear mixed models, with its principal advantage being that it produces less biased estimates of the variance components. However, the concept of REML does not immediately generalize to the setting of non-normally distributed responses, and it is not always clear the extent to which, either asymptotically or in finite samples, such generalizations reduce the bias of variance component estimates compared to standard unrestricted maximum likelihood estimation. 
In this article, we review various attempts that have been made over the past four decades to extend REML estimation in generalized linear mixed models. We establish four major classes of approaches, namely approximate linearization, integrated likelihood, modified profile likelihoods, and direct bias correction of the score function, and show that while these four classes may have differing motivations and derivations, they often arrive at a similar if not the same REML estimate. We compare the finite sample performance of these four classes, along with methods for REML estimation in hierarchical generalized linear models, through a numerical study involving binary and count data, with results demonstrating that all approaches perform similarly well reducing the finite sample size bias of variance components. Overall, we believe REML estimation should more widely adopted by practitioners using generalized linear mixed models, and that the exact choice of which REML approach to use should, at this point in time, be driven by software availability and ease of implementation.
\end{abstract}


\end{frontmatter}

\section{Introduction}\label{sec:Intro}

Restricted maximum likelihood (REML) estimation is a widely known and commonly used method for fitting linear mixed models (LMMs). The principal advantage of REML estimation comes from the fact that, for LMMs, standard unrestricted maximum likelihood estimation produces estimators of the variance components which are biased towards zero because it does not account for the degrees of freedom lost by estimating fixed effect coefficients. REML estimation corrects for the degrees of freedom lost, leading to less biased estimators of the variance components and more generally of the random effects covariance matrix. 

Historically, REML estimation is often attributed to the seminal contributions of \cite{patterson1971recovery,patterson1974maximum}, although prior to these, developments had been made for some balanced analysis of variance (ANOVA) models \citep[][]{anderson1952statistical,russell1958one}, and then `all' balanced ANOVA models by \cite{thompson1962problem}. Following the work of \citet{patterson1971recovery,patterson1974maximum}, the restricted likelihood function was defined as the marginal likelihood of a set of residual contrasts, from which the idea of REML estimation for LMMs arises. 
\cite{harville1974bayesian} formulated an alternative derivation of REML based on integrated likelihood using Bayesian arguments, while \cite{cooper1977note} provided another derivation focusing on the autoregressive-moving average model. 
\cite{laird1982random} showed that in balanced ANOVA models, REML estimation is equivalent to maximizing the likelihood obtained by integrating out the fixed effects using a diffuse prior. Another derivation of REML for LMMs comes from \cite{verbyla1990conditional}, who presented a conditional derivation related to that of \cite{harville1974bayesian}; see Section \ref{subsec:reml_LMMs}. 
For LMMs specifically, REML was further shown to be a special case of approximate conditional likelihood \citep{cox1987parameter}, adjusted profile likelihood \citep{mccullagh1990simple} and modified profile likelihood \citep[Chapter 9,][]{severini2000likelihood}. Ultimately, all these derivations of REML for LMMs arrive at the same objective function, and hence the same estimators of the random effects covariance matrix. 

Despite the extensive literature on REML estimation in LMMs, it is not immediately obvious how to define REML outside this class of models, since zero-mean residual contrasts generally do not exist in nonlinear models \citep{smyth1996conditional}. One consequence, and the main focus of this review, is that there have been many attempts to extend REML estimation to generalized linear mixed models (GLMMs) with non-normally distributed responses and a non-identity link function (see, 
\citeauthor{mcculloch2004generalized}, \citeyear{mcculloch2004generalized},
for an overview of GLMMs). 
Unlike their linear counterparts where the same restricted likelihood function is ultimately reached, with GLMMs different formulations of REML can potentially lead to different objective functions and thus seemingly different REML estimators exhibiting varying empirical behaviour. 

REML estimation for GLMMs and non-normally distributed response models remains a subject of active research (see, e.g., \citeauthor{tawiah2019multilevel}, \citeyear{tawiah2019multilevel}, 
\citeauthor{mcneish2019poisson}, \citeyear{mcneish2019poisson}, 
\citeauthor{elff2021multilevel}, 
\citeyear{elff2021multilevel}, and 
\citeauthor{han2024enhanced}, \citeyear{han2024enhanced}), 
while REML methods continue to be applied in various contexts such as model selection \citep{verbyla2019note}, bias adjustment for the asymptotic covariance of LMM and GLMM parameters (e.g., \citeauthor{kenward1997small}, \citeyear{kenward1997small}, and \citeauthor{watson2023generalised}, \citeyear{watson2023generalised}), meta-analysis (e.g., \citeauthor{bakbergenuly2018meta}, \citeyear{bakbergenuly2018meta}, and \citeauthor{lin2020meta}, \citeyear{lin2020meta}), or factor analysis (e.g., \citeauthor{noh2019hierarchical}, \citeyear{noh2019hierarchical}). More generally, REML has found widespread adoption among practitioners and applied scientists in data analyses involving GLMMs thanks to the availability of software such as the package \textsf{glmmTMB} \citep{brooks2017glmmtmb} in \textsf{R} \citep{R2024} and the \textsf{GLIMMIX} procedure in \textsf{SAS} \citep{SAS2015}. In genome-wide association studies, the usage of REML, e.g., for estimation of genetic correlations, has also become popular thanks to procedures such as \textsf{BOLT-REML} \citep{loh2015contrasting} available in the software package \textsf{BOLT-LMM} \citep{loh2015efficient}.

In this article, we aim to provide an overview of the various REML estimation methods that have been proposed for GLMMs. We categorize these methods into four main classes of REML approaches: approximate linearization (the largest category), integrated likelihood, modified profile likelihood and direct bias correction of the score function methods.  
We also review REML estimation for the class of hierarchical generalized linear models (HGLMs), of which GLMMs can be considered a particular class.
A central result we find is that while these different approaches to REML estimation may have differing motivations and algorithmic implementations, they often arrive at a similar if not the same estimate.
A simulation study comparing several currently available methods for REML estimation in \textsf{R} for GLMMs with binary and count data reveals that most REML methods perform similarly well in reducing the finite sample bias of the variance components, offering noticeable bias reductions, especially when the number of fixed effects is large relative to the number of random effects.

The present literature on GLMMs lacks a comprehensive and up-to-date review of the various approaches for REML estimation. Partial summaries of REML procedures have been presented in the past, but these either focus on a particular response type such as binary GLMMs \citep{reichert1993variance, meza2009estimation}, or are tailored to particular cases such as HGLMs \citep{lee2003extended,noh2007reml}. We seek to fill this gap and provide a overarching review with a particular emphasis on methods that are applicable to a broad range of mixed models with non-Gaussian distributed responses. 

Our overall conclusions and take-home messages from this review can be summarized as follows. First, we believe the extension of REML to GLMMs should be more widely adopted by practitioners, given our numerical results clearly show that it is worthwhile when the goal is to reduce bias in variance component estimation in GLMMs. 
Second, given the current theoretical research on REML for GLMMs is rather piecemeal and more motivational than rigorous, we encourage more precise asymptotic investigations into the various flavors of REML estimation for GLMMs, for instance, deriving under what conditions it can remove the leading bias term in the estimation of variance components. 
Finally, given the above two points, we argue that at this current point in time the choice of which REML implementation to use in mixed model analysis should be driven by practical constraints such as software availability and ease of implementation more than theory.

In Section \ref{sec:likelihoodInference} we define GLMMs and one approach to constructing REML estimation for LMMs. We then categorize the different methods for REML estimation in GLMMs into four main approaches (based on approximate linearization, integrated likelihood, modified profile likelihood or direct bias adjustment for the profile score function) and describe each approach in its own section (Sections \ref{sec:approxLinearization}--\ref{sec:biasCorr}). Methods developed specifically for the class of HGLMs are reviewed in Section \ref{sec:hglms}. Section \ref{sec:simStudy} provides a numerical study assessing the performance of some REML estimation methods currently available and/or easily implementable in \textsf{R}. Finally, Section \ref{sec:lastSection} summarizes our take-home messages and offers some avenues for future research. As a final introductory remark, we note the first letter of the REML acronym has been used to refer to both ``restricted" and ``residual" maximum likelihood estimation. Historically, these two terms have been employed interchangeably, although in this paper we will use the term ``restricted" as it has found more widespread use in the literature. 

\section{Generalized Linear Mixed Models} \label{sec:likelihoodInference} 
Generalized linear mixed models, which encompass the special case of linear mixed models with normally distributed responses and an identity link, express the mean of the response distribution as a function of fixed and random effects. 
A general formulation of GLMMs is given by \citet{hui2021use} among others. For a vector $\by=(y_1,\ldots,y_n)^\top$ of $n$ responses, the corresponding conditional mean vector $\bmu=(\mu_1,\ldots,\mu_n)^\top$ is modeled as
\begin{equation}
\begin{array}{c}
    \by\vert\bu \sim \prod_{i=1}^n p(y_i\vert \bu;\bbeta, \phi) \\[1.5ex] 
    g(\bmu) = \bdeta = \bX\bbeta + \bZ\bu, \\[1.5ex]
    \bu \sim p(\bu;\btheta) = N\{\bzero, \bD(\btheta)\},
\end{array}
    \label{eq:baseGLMMmodel}
\end{equation}
where $g(\cdot)$ is a known link function applied element-wise to $\bmu$, $\bdeta=(\eta_1,\ldots,\eta_n)^\top$ denotes the linear predictor vector, $\bX$ and $\bZ$ denote $n \times p$ and $n \times q$ matrices comprising $p$ fixed and $q$ random effect covariates, respectively, and $\bbeta$ and $\bu$ denote the corresponding fixed and random effect coefficients. Conditional on the random effects, the $y_i$'s are assumed to be independent observations from an exponential family of distributions $p(y_i\vert \bu;\bbeta, \phi)$ with mean $\mbox{E}(y_i\vert\bu) = \mu_i$ and dispersion parameter $\phi > 0$; the latter appears in the conditional variance of the distribution, $\mbox{Var}(y_i\vert \bu) = \phi V(\mu_i)$.

Linear mixed models arise as a special case of \eqref{eq:baseGLMMmodel} when $p(y_i\vert \bu;\bbeta, \phi)$ is a normal distribution, $g(\bmu) = \bmu$ is set to the identity link, and $\phi$ is the error variance such that $V(\mu) = 1$. Two prominent non-normal cases of GLMMs are binomial GLMMs where $p(y_i\vert \bu; \bbeta, \phi)$ is the binomial distribution, $g(\mu) = \log\{\mu/(1-\mu)\}$ is the canonical logit link, $\phi = 1$ is known and $V(\mu) = \mu(1-\mu)$, and Poisson GLMMs where $p(y_i\vert \bu;\bbeta, \phi)$ is the Poisson distribution, $g(\mu) = \log(\mu)$ is the canonical log link, $\phi = 1$ is known and $V(\mu) = \mu$.
The above formulation can be further expanded to involve more than one dispersion parameter, e.g., for Tweedie responses, and the REML methods reviewed in this article can be modified to accommodate such settings. However, for ease of presentation we will focus on the above formulation and refer the reader to \cite{mccullagh1989generalized} and \cite{wood2017generalized} for general overviews on exponential families of distributions for regression modeling.

We assume the random effects in \eqref{eq:baseGLMMmodel} are drawn from a multivariate normal distribution with a mean vector of zeros, and a random effects covariance matrix $\mbox{Cov}(\bu) = \bD(\btheta)$ which depends on a set of parameters $\btheta$. While it is possible to consider non-normally distributed random effects (see Section \ref{sec:hglms} for a discussion on REML estimation in HGLMs), we restrict the focus of this review largely to the normal random effects case given this is the most prevalent assumption in practice \citep{breslow1993approximate, mcculloch2004generalized,stroup2012generalized}. Note $\bD(\btheta)$ can and often does possess a large amount of structure, e.g., it may have a block-diagonal or fully diagonal structure in the case of independent cluster GLMMs. The diagonal elements of $\bD(\btheta)$ are often referred to as variance components, although in this article we will use the terms ``variance components" and ``random effects covariance matrix" interchangeably to mean $\btheta$ or $\bD(\btheta)$.
 
Model \eqref{eq:baseGLMMmodel} is broad enough to accommodate many applications of GLMMs. For instance, for independent clustered, nested or multi-level GLMMs, the columns of $\bZ$ can be rearranged so that $\bZ = [\bZ_1 \ldots \bZ_K]$, with all the submatrices $\bZ_k$, $k = 1,\ldots,K$, being block-diagonal in form, while $\bD(\btheta)$ is also written in a block-diagonal form. Crossed-effect GLMMs are a case where the submatrices $\bZ_k$ are not simultaneously block-diagonal. In this article, we will as much as possible utilize this broad formulation of GLMMs to keep the review general, with specific cases used as necessary, e.g., when particular papers develop REML estimation for a specific case of \eqref{eq:baseGLMMmodel}, and also in our simulation study. 

The marginal log-likelihood function for the GLMM formulated above is given by
\begin{equation}
    \ell_M(\bbeta,\phi,\btheta;\by) = \log \left\{\int p(\by\vert\bu;\bbeta,\phi)p(\bu;\btheta)d\bu\right\}.
    \label{eq:GLMMmargLogLik}
\end{equation}
Standard unrestricted maximum likelihood estimation is based on optimizing equation \eqref{eq:GLMMmargLogLik} or some variation of this such as approximated marginal log-likelihoods obtained via Laplace, quadrature, variational, and Monte-Carlo approximations; see for example, \cite{breslow1993approximate}, \cite{shun1995laplace}, \cite{mcculloch1997maximum}, \cite{ormerod2012gaussian}, \cite{hall2020fast}, and \cite{hui2021use} and references therein. 
Often central to fitting GLMMs is the joint log-likelihood function of the responses and random effects 
\begin{align}
\begin{split}
    \ell_J(\bbeta,\phi,\btheta;\by,\bu) = \log\{p(\by\vert\bu;\bbeta,\phi)\}\\ 
    + \log\{p(\bu;\btheta)\};
\end{split}
    \label{eq:GLMMjointLogLik}
\end{align}
in particular the Laplace approximation is implemented around the maximizer of the joint likelihood function \citep{breslow1993approximate,shun1995laplace}.

\subsection{REML Estimation for Linear Mixed Models} \label{subsec:reml_LMMs}

For LMMs, standard unrestricted maximum likelihood estimation based on equation \eqref{eq:GLMMmargLogLik} can proceed in a number of ways. For estimating the variance components $\btheta$ and the dispersion parameter $\phi$,
one approach is to maximize the profile log-likelihood function $\ell_P(\phi,\btheta;\by) = \ell_M(\widehat{\bbeta}_{(\phi,\btheta)},\phi,\btheta;\by)$, obtained by replacing the fixed effect coefficients with their maximum likelihood estimates for given fixed values of $\phi$ and $\btheta$, $\widehat{\bbeta}_{(\phi,\btheta)}$. Regardless of the exact procedure, maximum likelihood estimation does not account for the degrees of freedom lost in estimating the fixed effect coefficients $\bbeta$. This results in biased estimators of the variance components and dispersion parameters. For instance, in independent cluster LMMs the standard maximum likelihood estimator of $\btheta$ can be (heavily) biased towards zero if the number of clusters is small relative to the number of fixed effects. 

To address the above problem, \citet{patterson1971recovery,patterson1974maximum} introduced the idea of REML estimation for LMMs, based on maximizing a restricted log-likelihood function. 
%
%
REML can be derived in a number of ways, and here we review the conditional derivation of \cite{verbyla1990conditional} for an LMM of the form 
\begin{equation}
\begin{array}{c}
    \by = \bX\bbeta + \bZ\bu + \bvarepsilon,\quad \bu\sim N(\bzero,\sigsqH\bG),\\[1.5ex]
    \bvarepsilon\sim N(\bzero,\sigsqH\rho \bR),
\end{array}
    \label{eq:LMMformulation}
\end{equation}
where $\bvarepsilon$ is an $n$-vector of error terms. This form can be written as a case of model \eqref{eq:baseGLMMmodel} with $\bD(\btheta) = \sigma^2 \bG$, $\btheta = (\sigma^2, \text{vech}(\bG)^\top)^\top$ and $\phi = \sigma^2\rho$, where $\text{vech}(\cdot)$ denotes the half-vectorization operator, and we estimate $\rho$ instead of the dispersion parameter $\phi$ directly. 
The formulation (\ref{eq:LMMformulation}) allows for residual correlations between the elements of $\by$ conditional on $\bu$, as quantified by the elements of $\bR$. The joint log-likelihood function of the LMM \eqref{eq:LMMformulation} is given by
\begin{align*}
    &\ell_{\text{LMM},J}\{\bbeta,\rho,\btheta,\text{vech}(\bR);\by,\bu\}=-\frac{1}{2}\left\{\log\det(\sigsqH\bG)\right\} \\
    & -\frac{1}{2}\left\{\log\det(\sigsqH\rho\bR)\right\} - \frac{1}{2\sigma^2}\bu^\top \bG^{-1} \bu\\
    &-\frac{1}{2\sigma^2\rho}(\by-\bX\bbeta-\bZ\bu)^\top \bR^{-1} (\by-\bX\bbeta-\bZ\bu),
\end{align*}
where terms constant with respect to model parameters are omitted. In the conditional derivation of the REML likelihood function, we seek to maximize the portion of the joint log-likelihood function not associated with the fixed effects coefficients. To this end, consider the transformation $(\by_1^\top,\by_2^\top)^\top=\bL^\top\by$, where $\bL=[\bL_1 \; \bL_2]$ is a non-singular matrix. The matrices $\bL_1$ and $\bL_2$ are of dimension $n\times p$ and $n\times(n-p)$, respectively, and satisfy $\bL_1^\top\bX = \bI_p$ with $\bI_p$ being the $p \times p$ diagonal matrix, and $\bL_2^\top\bX=\bO_{n-p,p}$ with $\bO_{n-p,p}$ being the $(n-p)\times p$ matrix of zeros. We can then straightforwardly show that the distribution of this transformed response vector is 
\begin{equation*}
	\left[\begin{array}{c} \by_1\\[1ex] \by_2 \end{array}\right]\sim N\left(\left[\begin{array}{c} \bbeta\\[1.5ex] \bzero_{n-p}\end{array}\right],\sigsqH\left[\begin{array}{cc} \bL_1^\top \bV \bL_1 & \bL_1^\top \bV \bL_2  \\[1ex] \bL_2^\top \bV \bL_1 & \bL_2^\top \bV \bL_2 \end{array}\right]\right),
\end{equation*}
where $\bV = \rho\bR+ \bZ \bG \bZ^\top$ and $\bzero_{n-p}$ is an $n-p$ vector of zeros. It follows that the joint log-likelihood $\ell_{\text{LMM},J}\{\bbeta,\rho,\btheta,\text{vech}(\bR);\by,\bu\}$ can be written as the sum of the two contributions: one from the conditional distribution of $\by_1$ given $\by_2$, and one from the marginal distribution of $\by_2$. That is, 
\begin{equation*}
\begin{array}{c}
     \by_1\,\vert\,\by_2 \sim N\big(\bbeta - \bL_1^\top\bV\bL_2(\bL_2^\top\bV\bL_2)^{-1}\by_2,\\[1ex]\qquad\qquad\sigsqH(\bX^\top\bV^{-1}\bX)^{-1}\big)\\[1.5ex]
     \mbox{and}\quad \by_2 \sim N\big(\bzero,\sigsqH\bL_2^\top\bV\bL_2\big).
\end{array}
\end{equation*}
It is the marginal distribution of $\by_2$ which forms the restricted log-likelihood function
\begin{align}
    \begin{split}
    &\ell_{\text{LMM},R}\{\rho,\btheta,\text{vech}(\bR);\by\} = -\frac{1}{2}\Big\{(n-p)\log(\sigsqH)\\
    &\quad +\log\det(\bL_2^\top\bV\bL_2)\Big\}-\frac{1}{2\sigma^2}\by_2^\top(\bL_2^\top\bV\bL_2)^{-1}\by_2 \\
    &= -\frac{1}{2}\left\{(n-p)\log(\sigsqH)+\log\det(\bV)\right\}\\
    &\quad- \frac{1}{2\sigma^2}\by^\top\bP\by -\frac{1}{2} \log\det(\bX^\top\bV^{-1}\bX), 
    \end{split}
    \label{eq:REMLloglik}
\end{align}
where $\bP = \bV^{-1}-\bV^{-1}\bX\big(\bX^\top\bV^{-1}\bX\big)^{-1}\bX^\top\bV^{-1}$ and constants are again omitted. Note expression \eqref{eq:REMLloglik} does not depend on the fixed effect coefficients $\bbeta$. The final term $-(1/2)\log\det(\bX^\top \bV^{-1}\bX)$ is sometimes known in the literature as the ``REML adjustment", and forms similar to this will appear throughout this review. Maximization of $\ell_{\text{LMM},R}\{\rho,\btheta,\text{vech}(\bR);\by\}$ leads to REML estimators of the random effects covariance matrix and dispersion parameter.
It can be shown that REML estimators have the properties of likelihood-based estimators, in that they are asymptotically normally distributed with covariance matrix based on the information matrix derived from $\ell_{\text{LMM},R}\{\phi,\btheta,\text{vech}(\bR);\by\}$, and assuming here the dimension of $\bbeta$ ($p$) does not grow with sample size ($n$); see \cite{verbyla1990conditional} and \cite{cressie1993asymptotic} for details.

The conditional derivation above represents one of several alternative derivations by which REML estimators can be derived for LMMs. Importantly, all formulations of REML for LMMs produce the same objective function as that given in equation \eqref{eq:REMLloglik}. By contrast, this variety of derivations has inspired a multitude of REML likelihood functions and estimators in GLMMs. We review these in the next five sections.

\section{Approximate Linearization \label{sec:approxLinearization}}

The greater part of the research on REML estimation for GLMMs uses an approximate linearization strategy. 
Such procedures involve replacing the GLMM by an approximate linear mixed model, after which various LMM results for constructing REML estimates can be adapted. We choose to review approximate linearization methods through three sets of representative, connected papers: \cite{schall1991estimation} and \cite{wolfinger1993generalized} who obtained REML estimates by applying results on an adjusted dependent variable, \citet{mcgilchrist1994estimation} whose computation of REML estimates is framed within an iterative scheme for calculating best linear unbiased predictors, and \cite{breslow1993approximate} and \cite{engel1994simple} who proposed quasi-likelihood estimation methods for GLMMs. 

\subsection{Linearization and Adjusted Dependent Variables}\label{subsec:schall}

\citet{schall1991estimation} considered a GLMM of the form 
\begin{equation*}
    \begin{array}{c}
    g(\bmu) = \bX\bbeta + \sum_{k=1}^K \bZ_k\bu_k,\\[1.5ex]
    \mbox{Cov}(\bu)=\bD(\btheta)=\diag(\sigma^2_1\bI_{q_1},\ldots,\sigma^2_K\bI_{q_K}),
    \end{array}
\end{equation*}
where $\btheta = (\sigma^2_1,\ldots,\sigma^2_K)^\top$ are the variance components. 
This model is a special case of \eqref{eq:baseGLMMmodel} where the random effects model matrix comprises a set of $k = 1,\ldots,K$ sparse submatrices $\bZ_k$ of dimension $n\times q_k$, with $q=\sum_{k=1}^K q_k$, while the corresponding random effects $\bu = (\bu_1^\top, \ldots, \bu^\top_K)^\top$ are structured such that the $q_k$ elements of $\bu_k$ are drawn independently from a random effects distribution with mean zero and variance component $\sigma^2_k$. 
This particular form of GLMM is often associated with nested/multi-level data where only random intercepts are included, as well as data from crossed-effect designs.
We remark that the original formulation in \citet{schall1991estimation} makes assumptions only on the expectation and variance of the random effects distribution, although the results shown below assume normality for $\bu$. More generally, the approach of \citet{schall1991estimation} which we describe below only applies to GLMMs with independent random effects, and strictly to cases where the random effects precision matrix is linear in the variance parameters \citep[see also Section 6.4 of][]{wood2017generalized}.

The method of \citet{schall1991estimation} uses an approximate linearization of the link function $g(y_i) \approx \eta_i + g'(\mu_i) (y_i - \mu_i)\triangleq \xi_i$, where $\bxi = (\xi_1,\ldots,\xi_n)^\top$ denotes a vector of adjusted dependent variables \citep[also known as working responses;][]{mccullagh1989generalized}.
Note this linearization is commonly used in the context of estimation for generalized linear and generalized additive models  \citep[via iterative reweighted least squares;][]{wood2017generalized}. Based on this adjusted dependent variable we can construct an approximate linearized GLMM as
\begin{equation}
    \bxi = \bX\bbeta + \bZ \bu + \be,\quad \mbox{E}(\be) = \bzero, \quad \mbox{Cov}(\be) = \bW^{-1}, 
    \label{eqn:linearizedGLMM}
\end{equation}
where $\bW^{-1}$ is a diagonal weight matrix with diagonal elements $\text{Var}\{g'(\mu_i) (y_i - \mu_i) \vert \bu \} = \phi V(\mu_i) \{g'(\mu_i)\}^2$.
Regarding the expressions in \eqref{eqn:linearizedGLMM} as an LMM with marginal mean and covariance $\mbox{E}(\bxi|\bu) = \bX\bbeta$ and $\mbox{Cov}(\bxi) = \bW^{-1} + \bZ \bD(\btheta) \bZ^\top$, we can then apply established methods for computing REML estimates in LMMs. In particular, \citet{schall1991estimation} proposed a two-step iterative procedure adapted from \citet{fellner1986robust, fellner1987sparse}, which itself is built upon the work of \citet{henderson1963selection} and \cite{harville1977maximum}. 
First, given current values of all the parameters $(\bbeta^\top,\phi,\btheta^\top)^\top$ and random effects $\bu$, update the adjusted dependent variable $\bxi^\dagger = \bX\bbeta + \bZ \bu$, and compute $\bD(\btheta)$ and $\bW$. Then, given these updated values, compute $\bbeta$ and $\bu$ by solving the mixed model equations of \citet{henderson1963selection}, 
\begin{align} 
\begin{split}
\bA \left[\begin{array}{c} \bbeta\\[1ex] \bu\end{array}\right] &= \left[\begin{array}{cc} \bX^\top \bW\bX & \bX^\top \bW\bZ \\[1ex] \bZ^\top \bW \bX & \bZ^\top \bW\bZ + \bD(\btheta)^{-1} \end{array}\right] \left[\begin{array}{c} \bbeta\\[1ex] \bu\end{array}\right]\\
&= \left[\begin{array}{c} \bX^\top \bW\bxi^\dagger\\[1ex] \bZ^\top \bW\bxi^\dagger \end{array}\right].
\end{split}
\label{eqn:mixedmodeleqns}
\end{align}
Finally, to update the REML estimates of the variance components, let 
\begin{align*}
    \bT&= \{\bZ^\top \bW \bZ \\
    &\,\,\quad- \bZ^\top \bW \bX (\bX^\top \bW \bX)^{-1} \bX^\top \bW \bZ + \bD(\btheta)^{-1}\}^{-1}
\end{align*}
denote the matrix formed by the last $q$ rows and $q$ columns of $\bA^{-1}$. 
Partition $\bT$ into blocks $\bT_{kk'}$ for $k,k' = 1,\ldots,K$ conformably with the diagonal structure of $\bD(\btheta)$, and update the elements of $\btheta$ via the fixed-point formula $\sigma^2_k \Leftarrow (\bu_k^\top\bu_k)/\{q_k - \sigma^{-2}_k\mbox{trace}(\bT_{kk})\}$.
If the dispersion parameter $\phi$ also requires estimation, a further step can be taken again using a REML-like estimator; see \citet{fellner1986robust} for more details. 

\citet{schall1991estimation} pointed out that the proposed approach to REML estimation produces the same estimates of the variance components as the algorithm proposed by \cite{stiratelli1984random} using the integrated likelihood approach, which we review in Section \ref{sec:integratedLikelihood}. It is also worth noticing that the method of \cite{schall1991estimation} can be obtained simply from the derivative of the Laplace-approximate marginal log-likelihood computed with respect to the fixed and random effects; see Section 6.4.1 of \citet{wood2017generalized} for further details as well as additional connections to the penalized quasi-likelihood approach reviewed in Section \ref{subsec:pql}.
In the special case of a binomial GLMM using a probit link and single set of random effects, i.e., $K = 1$, the approximate linearization approach of \citet{schall1991estimation} also corresponds to the joint maximization method of \citet{gilmour1985analysis}. 

\citet{wolfinger1993laplace} extended the work of \citet{schall1991estimation} to construct REML estimates for nonlinear mixed models,
although \citeauthor{wolfinger1993laplace}'s approach applies the Laplace approximation to the marginal log-likelihood function of the nonlinear mixed model along with an improved first-order approximation to the nonlinear function of \citet{lindstrom1990nonlinear}; a REML adjustment akin to the final log-determinant term of \eqref{eq:REMLloglik} for estimating the variance components is then incorporated. Of more relevance here is the work of \citet{wolfinger1993generalized}, who studied GLMMs in the form given by \eqref{eq:baseGLMMmodel}, i.e., allowing for more general structures in $\bD(\btheta)$ compared to \citet{schall1991estimation}. Specifically, 
\citet{wolfinger1993generalized} arrived at the approximate linearized GLMM \eqref{eqn:linearizedGLMM}, but then adapted the work of \cite{lindstrom1990nonlinear} and explicitly assumed a multivariate normal distribution for $\be$. That is, their strategy is to directly specify $\bxi \sim N(\bX\bbeta + \bZ \bu, \bW^{-1})$ and then, analogously to Section \ref{subsec:reml_LMMs} except that $\phi$ is also profiled, obtain the restricted log-likelihood function 
\begin{align}
\begin{split}
    \ell_{\text{lin}, R}(\btheta;\by) &= -\frac{1}{2}\log\det(\bV) - \frac{n-p}{2}\log\br^\top\bV^{-1}\br\\
    &\quad- \frac{1}{2} \log\det(\bX^\top\bV^{-1}\bX),
\end{split}
\label{eqn:wolfinger_REML}
\end{align}
where $\br = \bxi - \bX (\bX^\top \bV^{-1} \bX)^{-1} \bX^\top \bV^{-1} \bxi$, $\bV = \bW^{-1} + \bZ \bD(\btheta)\bZ^\top$, and constants with respect to the model parameters are omitted. Note the $\bV$ matrix here is also a function of $\bbeta$ and $\bu$ through $\bW$, unlike in LMMs. \citet{wolfinger1993generalized} then iterated between solving the mixed model equations in \eqref{eqn:mixedmodeleqns} to update $\bbeta$ and $\bu$ plus using a profile likelihood estimate of $\phi$, and maximizing $\ell_{\text{lin}, R}(\btheta)$ to update $\btheta$. More generally, equation \eqref{eqn:wolfinger_REML} may be regarded as a GLMM analogue of the restricted log-likelihood function in \eqref{eq:REMLloglik}, achieved through approximate linearization, where the term $-(1/2)\log\det(\bX^\top\bV^{-1}\bX)$ is a REML adjustment. 

\citet{wolfinger1993generalized} noted their estimation procedure bears strong similarities to, and in some cases produces the same REML estimates as, the quasi-likelihood procedures of \cite{breslow1993approximate} which we review in Section \ref{subsec:pql}. Indeed, aside from whether $\phi$ is estimated versus assumed known, \citet{breslow1993approximate} derived their estimation procedure using a Laplace-approximated marginal quasi-likelihood function, while  \citet{wolfinger1993generalized} applied a Gaussian approximation directly to the adjusted dependent variables. \citet{wolfinger1993generalized} also noted that, in the special case of the GLMM studied in \citet{schall1991estimation}, their method produces the same REML estimates of $\sigma^2_k$ as in \citet{schall1991estimation}, with only minor differences in the fitting algorithm due to the use of the Newton-Raphson algorithm to maximize \eqref{eqn:wolfinger_REML}. 

\subsection{Linearization and the BLUP approach}
  
\citet{mcgilchrist1994estimation} developed a class of approximate linearization methods based on the method of best linear unbiased predictions (BLUPs) used to fit general mixed-effects models, and exploited the connection between BLUPs and REML for LMMs, as developed by \citet{henderson1963selection,henderson1975best}.
Consider the model $g(\bmu) = \bX\bbeta + \sum_{k=1}^K \bZ_k\bu_k$ and $\mbox{Cov}(\bu)=\bD(\btheta) =\diag(\sigma^2_1\bA_{1},\ldots,\sigma^2_K\bA_{K})$. This is similar to the model in \citet{schall1991estimation}, with the addition of known constant matrices $\bA_1, \ldots, \bA_{K}$ that do not have to be diagonal in structure. From \eqref{eq:GLMMjointLogLik}, the joint log-likelihood of the GLMM is
\begin{align*}
&\ell_J(\bbeta,\phi,\btheta;\by,\bu) = \log\{p(\by\vert\bu;\bbeta,\phi)\}\\
&\quad-\frac{1}{2} \sum_{k=1}^K \left\{q_k\log(\sigma^2_k) + \frac{1}{\sigma^2_k} \bu^\top_k \bA_k^{-1} \bu_k\right\}\triangleq \ell_1 + \ell_2,
\end{align*}
where constants with respect to model parameters are omitted. Next, let $\bH = -\nabla^2_{\bbeta,\bu}\ell_1$ denote the observed information matrix for $\bbeta$ and $\bu$; the expected information matrix could also be used, although this is anticipated to make little difference in practice.
Following arguments provided by \cite{mcgilchrist1991restricted} for LMMs, 
\citet{mcgilchrist1994estimation} then proposed replacing $\ell_1$ by a quadratic approximation 
around the estimators
$(\tilde{\bbeta}^\top,\tilde{\bu}^\top)^\top = \arg\max_{\bbeta, \bu} \ell_1$, 
\begin{align*}
    \ell^*_1 &= -\frac{1}{2}\left[\begin{array}{c} \bbeta - \tilde{\bbeta}\\[1ex] \bu - \tilde{\bu}\end{array}\right]^\top \bH \left[\begin{array}{c} \bbeta - \tilde{\bbeta}\\[1ex] \bu - \tilde{\bu}\end{array}\right] 
    \\
    &=-\frac{1}{2}(\by^* - \bX\bbeta - \bZ\bu)^\top\bW(\by^*-\bX\bbeta-\bZ\bu),
\end{align*}
where $\by^* = \bX\tilde{\bbeta} + \bZ\tilde{\bu}$ is an estimated response vector, and the second equality follows from studying the form of $\bH$ for the exponential families of distributions, with $\bW$ being the diagonal weight matrix defined below model \eqref{eqn:linearizedGLMM}. Because the quadratic approximation is constructed only from $\ell_1$, then strictly speaking $\bbeta$ and $\bu$ are generally not identifiable as one or more of the columns in $\bX$ and $\bZ$ are collinear with one other. However, since the final form of $\ell^*_1$ only requires the estimated response $\by^*$, and $\bW$ is a diagonal matrix that can be inverted, then the estimation algorithm based on this quadratic approximation itself encounters no issues. 

Viewing the approximate joint log-likelihood function $\ell^*_J(\bbeta,\phi,\btheta;\by,\bu) = \ell^*_1 + \ell_2$ as one corresponding to that of an approximate linearized GLMM, \citet{mcgilchrist1994estimation} then applied a standard procedure for estimating LMMs as follows. Given $\btheta$ and $\phi$, we calculate the BLUP estimators of $\bbeta$ and $\bu$ by maximizing $\ell^*_J(\bbeta,\phi,\btheta;\by,\bu)$. This can be shown to be equivalent to solving the mixed model equations \eqref{eqn:mixedmodeleqns}, with $\bxi^\dagger$ replaced by $\by^*$. Then, given $\bbeta$ and $\bu$, we compute updates for the REML estimates of $\sigma^2_k$ similar to those of \citet{schall1991estimation}: for $k = 1,\ldots,K$, we use the fixed-point update $\sigma^2_k \Leftarrow (\bu_k^\top \bA^{-1}_k \bu_k) / \{q_k - \sigma^{-2}_k\text{trace}(\bA^{-1}_k\bT_{kk})\}$. If $\phi$ requires estimation, then this can also be updated using a REML-like estimator. 

Under the same choice of mean model, response distribution, and link function, the approximate linearization methods of \citet{mcgilchrist1994estimation} and \citet{schall1991estimation} are equivalent and produce the same REML estimates. This equivalence was noted by \citet{mcgilchrist1994estimation}, who commented that the primary difference between the two methods lies in arguments by which the approximation linearization is obtained: \citet{schall1991estimation} and \citet{wolfinger1993generalized} considered linearization of the link function directly, while the linearization of \citet{mcgilchrist1994estimation} arises through approximating the first part of the joint log-likelihood function. In a small simulation study with binomial GLMMs, \citet{mcgilchrist1994estimation} showed their REML estimates exhibit less bias than a standard unrestricted maximum likelihood estimator, which they also derive, but may still present substantial bias if the true value of the variance component is large. On the other hand, \citet{mcgilchrist1994estimation} argued that the BLUP framework aims to target a broader class of mixed model applications. This is exemplified later by \cite{mcgilchrist1995derivation} who used the BLUP framework for REML estimation in a logistic GLMM with an AR(1) random component,
\citet{saei1997random} who used it in ordinal response mixed models, and \cite{yau1998ml} and \citet{yau2001multilevel} who applied it to frailty models in survival analysis. The BLUP approach to calculating REML estimates has been further extended by \citet{lee1996hierarchical} to HGLMs; see Section \ref{sec:hglms}.

\subsection{Linearization and Quasi-Likelihood Estimation} \label{subsec:pql}

Quasi-likelihood methods exploit the relationship between the mean and variance of the responses to construct an objective function, without requiring the specification of a full distribution in the model
\citep[][Chapter 9]{mccullagh1989generalized}. In the setting of GLMMs, this was adopted by \cite{breslow1993approximate} who proposed two quasi-likelihood methods, and subsequently developed REML estimation, for fitting GLMMs; see also \citet{green1987penalized} and \citet{goldstein1991nonlinear} for the same approach in the context of semi-parametric regression models and nonlinear multi-level mixed models, respectively.

Consider the mean model in \eqref{eq:baseGLMMmodel}, but instead of an exponential family of distributions, we assume only conditional first- and second-order moment forms for the response. That is, $g(\bmu) = \bdeta = \bX\bbeta + \bZ\bu$ and $\mbox{Var}(y_i\vert \bu) = \phi a_i V(\mu_i)$ for $i=1,\ldots,n$, where $a_i$ is a constant that may vary across observations, e.g., the number of trials in the case of binomial responses, and $V(\mu)$ is an assumed variance function. If the random effects are normally distributed with zero means and covariance matrix $\bD(\btheta)$, then we can define the quasi-likelihood function for this GLMM as 
\begin{align*}
    &\ell_{\text{quasi},M}(\bbeta,\phi,\btheta;\by) = -\frac{1}{2}\log\det\{\bD(\btheta)\}\\
    &\qquad\qquad+ \log \left[\int \exp\{-\kappa(\bbeta,\phi,\btheta;\by,\bu)\} d\bu\right],
\end{align*}
where 
$$\kappa(\bbeta,\phi,\btheta;\by,\bu) = \frac{1}{2\phi}\sum_{i=1}^n d_i(y_i,\mu_i) + \frac{1}{2}\bu^\top\bD(\btheta)^{-1}\bu,$$ 
and $d_i(y,\mu) = -2\int_{y}^{\mu} (y-t)/\{a_i V(t)\}dt$ can be interpreted as a conditional deviance measure. Next, we apply a Laplace approximation \citep{tierney1986accurate} to the integral in $\ell_{\text{quasi},M}(\bbeta,\phi,\btheta;\by)$, based on a quadratic expansion around the minimum of $\kappa(\bbeta,\phi,\btheta;\by,\bu)$ as a function of $\bu$. 
Following some algebraic manipulation, and omitting 
one log-determinant term in the Laplace approximation under the assumption that it and the so-called ``GLM iterative weights" vary slowly as functions of the model parameters, \citet{breslow1993approximate} arrived at the penalized quasi-likelihood (PQL) function for fitting GLMMs
\begin{align} 
\label{eqn:PQL}
\begin{split}
    \ell_{\text{PQL},M}(\bbeta;\by,\bu) &= -\frac{1}{2\phi}\sum_{i=1}^n d(y_i,\mu_i)\\
    &\quad- \frac{1}{2}\bu^\top\bD(\btheta)^{-1}\bu.
\end{split}
\end{align}
We make two key remarks about expression \eqref{eqn:PQL}. First, $\ell_{\text{PQL},M}(\bbeta;\by,\bu)$ is closely connected to the joint log-likelihood function in \eqref{eq:GLMMjointLogLik}. If the deviance measure $d_i(y,\mu)$ corresponds exactly to a case from an exponential family of distributions, then $\ell_{\text{PQL},M}(\bbeta;\by,\bu)$ is the same as taking only the terms in $\ell_J(\bbeta,\phi,\btheta;\by,\bu)$ which are a function of fixed effect coefficients, random effects and dispersion parameters. As such, for a given value of $\btheta$ and $\phi$, maximizing the PQL function is the same as solving the mixed model equations given by \eqref{eqn:mixedmodeleqns} (cf. \citeauthor{wolfinger1993generalized}, \citeyear{wolfinger1993generalized} and \citeauthor{mcgilchrist1994estimation}, \citeyear{mcgilchrist1994estimation}). 

Instead of solving the mixed model equations though, \citet{breslow1993approximate} suggested using a Fisher scoring algorithm to perform the optimization for $\bbeta$ and $\bu$, which is algorithmically different although the resulting updates are the same. 
A related second point is that, by construction, equation \eqref{eqn:PQL} does not provide updates of the variance components and dispersion parameter, and we need to alternate between maximizing $\ell_{\text{PQL},M}(\bbeta;\by,\bu)$ and optimizing a second objective function to update $\btheta$ and $\phi$.  
For the latter, \citet{breslow1993approximate} used an approximate linearization procedure to construct the adjusted dependent variable $\xi_i = \eta_i + g'(\mu_i) (y_i - \mu_i)$, along with an accompanying $n\times n$ diagonal weights matrix $\bW^{-1}$ of weights with elements $\phi a_i V(\mu_i)\{g'(\mu_i)\}^2$, $ i=1,\ldots,n$. Then treating the adjusted dependent variables as normally distributed \citep[see also][Section 3.4.2]{wood2017generalized} and adapting the developments of \citet{harville1977maximum} for LMMs, given values of $\bbeta$ and $\bu$, the REML profile quasi-likelihood is formulated as
\begin{align}
\label{eqn:PQL_theta}
\begin{split}
    &\ell_{\text{PQL}, R}(\btheta;\by) = -\frac{1}{2}\log\det(\bV)\\
    &\qquad\qquad\qquad- \frac{1}{2}(\bxi^\dagger - \bX\bbeta)^\top\bV^{-1} (\bxi^\dagger -\bX\bbeta)\\
    &\qquad\qquad\qquad- \frac{1}{2}  \log\det(\bX^\top\bV^{-1}\bX),
\end{split}
\end{align}
where $\bxi^\dagger = \bX\bbeta + \bZ\bu$, $\bV = \bW^{-1} + \bZ\bD(\btheta)\bZ^\top$. Iterating between equations \eqref{eqn:PQL} and \eqref{eqn:PQL_theta} produces PQL estimates of the GLMM, including a set of REML estimates for the variance components. 
Importantly, the final term in equation \eqref{eqn:PQL_theta} takes the same form of REML adjustment as in \eqref{eqn:wolfinger_REML} used by \citet{wolfinger1993generalized}. Equation \eqref{eqn:PQL_theta} also corresponds to a case of the profile log-likelihood correction derived by \citet{cox1987parameter} for linear models, which was justified if $\bbeta$ and $\btheta$ are orthogonal and the information matrix for the estimator of the former is given exactly by $\bX^\top\bV^{-1}\bX$. While these conditions do not hold precisely for GLMMs, $\ell_{\text{PQL}, R}(\btheta,\phi;\by)$ is still regarded as a profile log-likelihood correction for an approximate linearized GLMM. 
We further note that \citet[][Section 3.4.2]{wood2017generalized} offered a reasonable but not rigorous large sample justification for using a linear mixed model based on adjusted dependent variables, and hence the REML profile quasi-likelihood approach in \eqref{eqn:PQL_theta}.

As remarked previously towards the end of Section \ref{subsec:schall}, it is not hard to see the close resemblance between PQL estimation and the pseudo-likelihood method of \citet{wolfinger1993generalized}: in many cases, the two approaches produce the same REML estimates of the variance components, and the primary difference lies in their motivation for their respective procedures. 
As noted by \citet{mcgilchrist1994estimation}, the PQL estimation approach also shares common elements with the BLUP method (see also \citeauthor{mcgilchrist1991restricted}, \citeyear{mcgilchrist1991restricted}, for LMMs), while the work of \cite{schall1991estimation} can also be recast within the PQL framework \citep[][Section 6.4.1]{wood2017generalized}. Furthermore, a close connection to the integrated likelihood approach reviewed in Section \ref{sec:integratedLikelihood} emerges from noticing that the derivative of $\ell_{\text{quasi},M}(\bbeta,\phi,\bu;\by)$ with respect to $\bbeta$ and $\bu$ leads to the same score equations derived by \cite{stiratelli1984random} for binary logistic GLMMs. 

As an aside, we remark that \citet{breslow1993approximate} also proposed a marginal quasi-likelihood (MQL) approach to fitting GLMMs. We provide some more details of this in the supplementary material, and only note here that REML profile quasi-likelihood estimation is also used to update $\btheta$ as part of this approach.

Since the work of \citet{breslow1993approximate}, it has been recognized that estimates of the model parameters obtained using PQL and MQL estimation can be severely biased, particularly when the response is discrete and/or the true elements of $\btheta$ are large. Several corrections have been proposed to reduce this bias (see, e.g., \citeauthor{breslow1995bias}, \citeyear{breslow1995bias}, \citeauthor{kuk1995asymptotically}, \citeyear{kuk1995asymptotically}, and \citeauthor{goldstein1996improved}, \citeyear{goldstein1996improved}), although
most corrections appear to be satisfactory only when the true variance components are relatively small. 
For instance, 
\cite{goldstein1996improved} reported serious biases for the REML estimates of variance components in their simulation study, and \cite{rodriguez1995assessment} showed that the second-order approach makes only a modest improvement to the original MQL procedure. 
A more subtle and relevant point here is that made by \citet{lin1996bias}, who eschewed correcting the REML estimator of the variance components in favor of correcting the standard, unrestricted profile quasi-likelihood estimator, where the latter is derived from an objective function similar to \eqref{eqn:PQL_theta} but which omits the final REML adjustment term. The reasoning for this is simple: for GLMMs, there is no theoretical guarantee the REML estimator of the variance components results in less biased estimators than the unrestricted quasi-likelihood estimator \citep[notwithstanding our note earlier about the reasonable asymptotic justification given in][]{wood2017generalized}. Therefore, if the aim is to make an adjustment for the asymptotic bias, then it is preferable to work with the less complex objective function that defines the unrestricted quasi-likelihood estimator; see also Section \ref{sec:lastSection} for our general remarks regarding the need for more theoretical investigation into REML estimation for GLMMs. 

Besides \cite{breslow1993approximate}, other attempts have been made to apply quasi-likelihood estimation to GLMMs, although these are confined to models with a single random effect. For instance, \citet{prentice1991estimating} considered discrete and continuous responses and mixtures thereof, while \citet{liang1992multivariate} examined binary and categorical responses. For binary data with a balanced design, \citet{reichert1993variance} showed both of these approaches lead to the same estimates of the variance components, in particular, they provide a REML-like estimate of the error variance and an ML-like estimate of the random effect variance. Finally, we acknowledge the work of \citet{engel1994simple}, who proposed to fit GLMMs by a combination of quasi-likelihood estimation performed using iterated least squares on an adjusted dependent variable, and minimum norm quadratic unbiased estimation \citep[MINQUE,][which is numerically equivalent to REML estimation]{mcculloch2004generalized} for the variance components. 
Although motivated differently, the method of \citet{engel1994simple} is equivalent to PQL estimation, including the use of the restricted profile quasi-likelihood function in \eqref{eqn:PQL_theta}, with the only differences between the two being minor variations in the order in which parameter estimates and predictions for the random effects are updated. 

Beyond the three flavors of approximate linearization examined above, there are a number of less well-known variations for fitting and computing REML estimates in GLMMs. We provide a summary of these variations in supplementary material S.3.

\section{Integrated Likelihood \label{sec:integratedLikelihood}}

Another well-established approach for constructing REML estimators in GLMMs is motivated by a Bayesian formulation of the model, where fixed effects are treated as random and assigned a prior distribution whose information tends to zero \citep{harville1974bayesian}; 
see \citet{berger1999integrated} for general overview of the use of integrated likelihood for eliminating nuisance parameters. Consider the GLMM in \eqref{eq:baseGLMMmodel} and suppose we adopt a flat prior for the fixed effect coefficients, $p(\bbeta) \propto 1$.
An integrated log-likelihood function is then obtained by marginalizing over both the fixed and random effects in the joint likelihood function, giving
\begin{align}
\begin{split}
&\ell_{\text{int},R}(\phi,\btheta;\by)\\
&\quad=\log \left[\int \exp\left\{\ell_J(\bbeta,\phi,\btheta;\by,\bu)\right\} p(\bbeta) d\bbeta d\bu\right] \\
&\quad= \log \left\{\int p(\by\vert\bu; \bbeta,\phi) p(\bu;\btheta) d\bbeta d\bu\right\}.
\end{split}
\label{eqn:integratedlikelihood}
\end{align}
The motivation for using the integrated likelihood approach as a form of REML estimation comes from the fact that for LMMs integrating over $\bbeta$ assuming a flat prior for these parameters leads exactly to the restricted log-likelihood function in equation \eqref{eq:REMLloglik}. This equivalence was first shown by \citet{laird1982random} in their REML analysis for balanced ANOVA models, although the extension to general LMMs is straightforward. 

With the above result in mind, a number of researchers have proposed using \eqref{eqn:integratedlikelihood} as a means of constructing REML estimators in GLMMs. One of the earliest comes from \citet{stiratelli1984random}, who considered binary GLMMs (with $\phi = 1$ known) for longitudinal data and proposed the following two-step iterative process. First, given $\btheta$, we update $(\bbeta^\top, \bu^\top)^\top$ by maximizing the posterior distribution of these parameters. Noting the flat prior on the fixed effect coefficients, $p(\bbeta, \bu\vert\phi,\btheta;\by) \propto \exp\{\ell_J(\bbeta,\phi,\btheta;\by,\bu)\} p(\bbeta) = \exp[\log\{p(\by\vert\bu;\bbeta,\phi)\} + \log\{p(\bu;\btheta)\}]$, from which we see that maximizing the posterior distribution is identical to both solving the mixed model equations in \eqref{eqn:mixedmodeleqns}, and maximizing the PQL function in \eqref{eqn:PQL} from the class of approximation linearization approaches.  
Next, denote the updated coefficient values as $(\widehat{\bbeta}^\top, \widehat{\bu}^\top)^\top$. Then 
to obtain REML estimates for $\btheta$, we maximize the integrated likelihood in \eqref{eqn:integratedlikelihood}. \citet{stiratelli1984random} achieved this via an Expectation-Maximization \citep[EM;][]{dempster1977maximum} algorithm. Here, the E-step is numerically intractable, so the conditional distribution of $(\bbeta^\top, \bu^\top)^\top$ given $\by$ is approximated by a multivariate normal distribution centered on the current estimate of the posterior mode and covariance matrix given by the inverse of the corresponding observed information matrix $\bOmega(\bbeta,\phi,\btheta;\by,\bu) = \left\{-\nabla^2_{\bbeta,\bu} \ell_J(\bbeta,\phi,\btheta;\by,\bu) \right\}^{-1}$, that is
\begin{equation}
p(\bbeta, \bu\vert\phi,\btheta;\by) \approx N\left(\left[\begin{array}{c} \widehat{\bbeta}\\[1ex] \widehat{\bu}\end{array}\right], \bOmega(\widehat{\bbeta},\phi,\btheta,\widehat{\bu};\by)\right).
\label{eqn:Stiratelli_approximatenormal}
\end{equation}
In making this approximation, the monotonically increasing behavior of the EM algorithm is no longer strictly guaranteed since the expectation is not performed with respect to the true posterior distribution. On the other hand, the approximation eases optimization of the integrated likelihood since, given $p(\bu;\btheta) = N\{\bzero, \bD(\btheta)\}$, and the E-step reduces to computing $\int \bu \bu^\top \, p(\bbeta, \bu\vert\phi,\btheta;\by) d\bu$, which has a closed form when employing \eqref{eqn:Stiratelli_approximatenormal}. In fact, as pointed out below \eqref{eqn:mixedmodeleqns}, when $\bD(\btheta) =\diag(\sigma^2_1\bI_{q_1},\ldots,\sigma^2_K\bI_{q_K})$
the REML estimates of the variance components produced from the integrated likelihood approach coincide with those of the approximate linearization method of \citet{schall1991estimation}; see Section 3 of \citet{schall1991estimation}, where this equivalence is stated in more detail.

Using an EM algorithm to maximize the integrated likelihood in \eqref{eqn:integratedlikelihood} was also explored by \citet{mcculloch1994maximum}, who developed a framework for REML (as well as unrestricted maximum likelihood) estimation of variance components for binary GLMMs. The framework is implemented using a threshold model with latent variables i.e., for binary responses we assume $y_i = \text{I}(w_i>0)$, $i=1,\ldots,n$, where $\text{I}(\cdot)$ denotes the indicator function and $w_i $ is a normal distributed random variable. An LMM is then specified for the $w_i$'s, with an EM algorithm used to optimize the resulting form for $\ell_{\text{int},R}(\phi,\btheta;\by)$.
\citet{mcculloch1994maximum} suggested the application of either numerical integration or Gibbs sampling \citep{gelfand1990sampling} for the E-step. 
Much later on, \citet{meza2009estimation} proposed a variation of this procedure for REML estimation of binary GLMMs using the probit link. Similar to \citet{mcculloch1994maximum}, a latent variable representation of the binary GLMM link is used. The fixed effects coefficients $\bbeta$ are (again) considered random and given a flat prior, but this time a stochastic approximated EM (SAEM) algorithm \citep{delyon1999convergence} is employed to update the subsequent integrated likelihood, which is reported to be more computationally efficient than a standard Monte Carlo EM algorithm, since samples are recycled from one iteration to the next. 


Although the EM algorithm and variations thereof are a standard approach to maximizing \eqref{eqn:integratedlikelihood}, in datasets where $n$, $p$ and/or $q$ are potentially quite large (remembering $\bbeta$ also needs to be integrated over), this approach can become computationally quite burdensome. As such, and with the rise of techniques such as automatic differentiation, a modern approach to REML estimation for GLMMs via integrated likelihood instead applies a (first-order) Laplace approximation directly to $\ell_{\text{int},R}(\phi,\btheta;\by)$. Following some straightforward algebraic manipulations,
the Laplace-approximated integrated log-likelihood is given by
%
%
\begin{align}
\begin{split}
    &\ell_{\text{int},R}(\phi,\btheta;\by) \approx \ell_J(\widehat{\bbeta},\phi,\btheta,\widehat{\bu};\by)\\
    &\qquad-\frac{1}{2}\log\det\{\bOmega(\widehat{\bbeta},\phi,\btheta,\widehat{\bu};\by)\} \\
    &\quad= 
    - \frac{1}{2} \log\det\{\bD(\btheta)\} - \frac{1}{2} \widehat{\bu}^\top\bD(\btheta)^{-1}\widehat{\bu}\\
    &\qquad-\frac{1}{2}\log\det\Big[\bZ^\top \Big\{\widehat{\bW}\\
    &\qquad- \widehat{\bW}\bX\left(\bX^\top \widehat{\bW} \bX\right)^{-1}\bX^\top\widehat{\bW}\Big\} \bZ\\
    &\qquad+ \bD(\btheta)^{-1} \Big]+\mbox{const},
    \label{eqn:integratedlikelihood_laplace}
\end{split}
\end{align}
where, given $\phi$ and $\btheta$, $$(\widehat{\bbeta}^\top,\widehat{\bu}^\top)^\top = \arg\max_{\bbeta, \bu} \ell_J(\bbeta,\phi,\btheta;\by,\bu),$$ 
$\widehat{\bW}$ is the inverse of the weight matrix defined below \eqref{eqn:linearizedGLMM} evaluated at $(\widehat{\bbeta}^\top,\widehat{\bu}^\top)^\top$, and `const' denotes terms that are constant with respect to $\phi$ and $\btheta$. Interestingly, the last term in  \eqref{eqn:integratedlikelihood_laplace} follows from recognizing that $\bOmega(\bbeta,\phi,\btheta;\by,\bu) \equiv \bA$, where $\bA$ was defined as part of the mixed model equations in \eqref{eqn:mixedmodeleqns}. Applying the block-inversion formula and focusing only on the terms that are a function of $\btheta$, we then obtain the stated result. Indeed, the term inside the final log determinant of \eqref{eqn:integratedlikelihood_laplace} is identical in form to the inverse of the $\bT$ matrix defined below equation \eqref{eqn:mixedmodeleqns} in the approximate linearization class of methods.

The Laplace-approximated integrated likelihood is available as part of the package \textsf{glmmTMB} in \textsf{R}, 
which uses a combination of the Laplace approximation and automatic differentiation to maximize \eqref{eqn:integratedlikelihood_laplace} and perform REML estimation for mixed models; we examine this method in Section \ref{sec:simStudy} as part of our simulation study.
The Laplace-approximated integrated likelihood approach to computing REML estimates has also been adopted in generalized additive models (see \citeauthor{wood2011fast}, \citeyear{wood2011fast} and \citeyear{wood2017generalized}, and the related package \textsf{mgcv} in \textsf{R}), which may be formulated as GLMMs with the smoothing parameters corresponding to inverse variance components. Numerical studies have shown that REML estimates of the smoothing parameters in such semiparametric regression models offer some improvement in estimation performance relative to generalized cross-validation or the Akaike information criterion, although differences with the standard unrestricted maximum likelihood estimation, i.e., using \eqref{eqn:integratedlikelihood} 
except that the fixed effects are not integrated over, have been generally shown to be minimal. 
Indeed, to our knowledge, there are no rigorous theoretical guarantees the (Laplace-approximated) integrated likelihood method for computing REML estimates explicitly reduces the leading bias term of the variance components in GLMMs relative to standard unrestricted maximum likelihood estimation. Despite this, in the simulation studies in Section \ref{sec:simStudy} we will show that the approach is effective at reducing finite sample bias in the GLMM setting.

\section{Modified Profile Likelihood \label{sec:profileLikelihood}}

The concept of a restricted log-likelihood function has well-established connections with both modified profile likelihood functions \citep{severini2000likelihood} and approximate conditional likelihood functions \citep{cox1987parameter}.
In particular, for LMMs both of these general likelihood approaches have been shown to produce REML estimators of the variance components (see \citeauthor{bellhouse1990equivalence}, \citeyear{bellhouse1990equivalence} and \citeauthor{cox1992note}, \citeyear{cox1992note}). With this in mind, \citet{bellio2011restricted} proposed yet another approach to REML estimation for GLMMs, by formulating a suitable modified profile likelihood which in the special case of LMMs leads to the restricted log-likelihood function \eqref{eq:REMLloglik}. Their modified profile likelihood function for REML estimation is defined as
\begin{align}
\begin{split}
    &\ell_{\text{MPL},R}(\btheta;\by)=\ell_P(\btheta;\by)\\
    &\qquad\qquad\quad+\frac{1}{2}\log\det\left\{\frac{\partial^2}{\partial\bbeta\partial\bbeta^\top}\ell_M(\bbeta,\btheta;\by)\right\}\\
    &\qquad\qquad\quad-\log\det\{\bC(\widehat{\bbeta},\widehat{\btheta};\widehat{\bbeta}_{\btheta},\btheta)\},
    \label{eq:ModProfLogLik}
\end{split}
\end{align}
where $\ell_P(\btheta;\by) = \ell_M(\widehat{\bbeta}_{\btheta},\btheta;\by)$ is the profile log-likelihood function obtained by replacing the unknown fixed effect coefficients with $\widehat{\bbeta}_{\btheta}$, i.e., the maximum likelihood estimates for given fixed values of $\btheta$, $(\widehat{\bbeta}^\top,\widehat{\btheta}^\top)^\top$ denote maximum likelihood estimates obtained from \eqref{eq:GLMMmargLogLik}, and $\bC(\widehat{\bbeta},\widehat{\btheta};\widehat{\bbeta}_{\btheta},\btheta)$ is a correction term. Among several asymptotically equivalent forms of this correction term, \cite{bellio2011restricted} use 
\begin{align*}               
&\bC(\bbeta_0,\btheta_0;\bbeta_1,\btheta_1)\\
&\quad=\mbox{Cov}_{\bbeta_0,\btheta_0}\left\{\frac{\partial}{\partial\bbeta}\ell_M(\bbeta_0,\btheta_0;\by),\frac{\partial}{\partial\bbeta}\ell_M(\bbeta_1,\btheta_1;\by)\right\},
\end{align*}
where $(\partial/\partial\bbeta)\ell_M(\bbeta_0,\btheta_0;\by)$ is the $\bbeta$ part of the marginal log-likelihood function computed at generic $\bbeta_0$ and $\btheta_0$ values, and $(\partial/\partial\bbeta)\ell_M(\bbeta_1,\btheta_1;\by)$ is defined similarly.
%
%
For the case of independent cluster GLMMs, i.e.,
the particular case of \eqref{eq:baseGLMMmodel} for $m$ independent groups and known $\phi$ (as assumed by \citeauthor{bellio2011restricted}, \citeyear{bellio2011restricted}, although we conjecture the method could be extended to the case when $\phi$ is also estimated), the correction term above can be approximated by
\begin{align*}
    &\widehat{\bC}(\widehat{\bbeta},\widehat{\btheta};\widehat{\bbeta}_{\btheta},\btheta)\\
    &\quad=\sum_{i=1}^m\left\{\frac{\partial}{\partial\bbeta}\ell_M(\widehat{\bbeta},\widehat{\btheta};\by_i)\right\}\left\{\frac{\partial}{\partial\bbeta}\ell_M(\widehat{\bbeta}_{\btheta},\btheta;\by_i)\right\}^\top,
\end{align*}
where $\ell_M(\bbeta,\btheta;\by_i)=\log\int p(\by_i\vert \bu_i;\bbeta)p(\bu_i;\btheta)d\bu_i$ denotes the marginal log-likelihood for the $i$th cluster with responses $\by_i=(y_{i1},\ldots,y_{in_i})^\top$. Among other properties,
it is possible to show that for independent cluster GLMMs the score and information bias of the modified profile log likelihood \eqref{eq:ModProfLogLik} vanishes asymptotically as $m\rightarrow\infty$, and that
the modified profile likelihood function is invariant to changes in the parametrization including those of variance components \citep[see also][Chapter 9]{severini2000likelihood}. On other other hand, neither \citet{bellio2011restricted} nor to our knowledge anyone else provided theoretical guarantees that $\ell_{\text{MPL},R}(\btheta;\by)$ reduces the finite sample bias of variance component estimates. Our simulations in Section \ref{sec:simStudy} empirically demonstrate that the modified profile likelihood method is reasonably effective at reducing finite sample bias when estimating variance components in GLMMs. 

One practical limitation of the modified profile likelihood approach to REML estimation is its application to more complex random effects settings. For instance, in GLMMs with crossed random effects, aside from the computational issue of higher dimensional integrals, the empirical approximation  $\bC(\widehat{\bbeta},\widehat{\btheta};\widehat{\bbeta}_{\btheta},\btheta)$ shown above does not straightforwardly apply due to a lack of any independence in the observations. Of course, the challenge of higher dimensional integrals with no analytic expressions affects all REML estimation approaches reviewed in this article, with the dimension of integrals increasing with the sample size in cases such as the crossed effects setting. On the other hand, the lack of independence in the observations is less of a limitation for methods such as integrated log-likelihood reviewed in the preceding section.

\section{Direct Bias Correction \label{sec:biasCorr}}

The fourth class of REML estimation methods we review comes from \citet{liao2002type}, and is based on the solution to a set of estimating equations derived by correcting the bias of the profile score function of the variance components. The idea of corrected profile score originates from \cite{mccullagh1990simple}, who utilized it as a way of constructing REML estimators for LMMs (again producing the same estimates as in Section \ref{subsec:reml_LMMs}) and it was then adopted by \cite{smyth1999adjusted} to estimate dispersion parameters in generalized linear models.
Assume $\phi$ is known, and let $(\widehat{\bbeta}^\top,\widehat{\btheta}^\top)^\top$ denote the standard maximum likelihood estimators from optimizing \eqref{eq:GLMMmargLogLik}.
Next, notice from the joint log-likelihood function \eqref{eq:GLMMjointLogLik} that for given $\bbeta$ the component corresponding to the variance components $\btheta$ is given by $-\log\det\{\bD(\btheta)\} - \tr[\{\bD(\btheta)\}^{-1}\bu\bu^\top]+\mbox{const}$. This then gives rise to $\widehat{\btheta}_{\bbeta}$, the maximum likelihood estimator of $\btheta$ for a known $\bbeta$. The approach of \citet{liao2002type} is then based upon comparing the score equation for $\widehat{\btheta}$ with that of $\widehat{\btheta}_{\bbeta}$, in order to obtain a REML estimator with less bias than the former. 

In detail, first define the function
\begin{equation*}
    h(\btheta;\bS)=-\frac{\partial}{\partial\btheta}\big(\log\vert\bD(\btheta)\vert+\tr[\{\bD(\btheta)\}^{-1}\bS]\big),
\end{equation*}
where $\bS$ is a matrix with the same dimensions as $\bD(\btheta)$ and is considered a constant when differentiated with respect to $\btheta$. It is straightforward to show the $r$-th element of $h(\btheta,\bS)$ is given by 
\begin{align*}
    &h(\btheta,\bS)_r \\
    &\quad=\tr\left[\{\bD(\btheta)\}^{-1}\{\bS-\bD(\btheta)\}\{\bD(\btheta)\}^{-1}\frac{\partial\bD(\btheta)}{\partial\theta_r}\right].
\end{align*}
Next, from a general identity for the EM algorithm \citep[p. 95,][]{mclachlan2007algorithm}, the score equation for $\widehat{\btheta}_{\bbeta}$ is given by
\begin{align}
\begin{split}
    &\frac{\partial}{\partial\btheta}\ell_M(\bbeta,\btheta;\by)\\
    &\quad =h\{\btheta;\mbox{E}_{\bu\vert\bbeta,\btheta;\by}(\bu\bu^\top\,\vert\,\bbeta,\btheta;\by)\}=\bzero,
\end{split}
    \label{eq:scoreEquation}
\end{align}
where 
the expectation of $\bu\bu^\top$ is with respect to the conditional distribution $p(\bu\vert\bbeta,\btheta;\by)\propto p(\by\vert\bu;\bbeta)p(\bu;\btheta)$.
Noting $\widehat{\btheta}$ maximizes the profile log-likelihood function $\ell_P(\btheta;\by) = \ell_M(\widehat{\bbeta}_{\btheta},\btheta;\by)$ (see also Section \ref{subsec:reml_LMMs}), then it follows from \eqref{eq:scoreEquation} that the profile score equation for $\widehat{\btheta}$ is given by
\begin{align}
\begin{split}
    &\frac{\partial}{\partial\btheta}\ell_P(\btheta;\by)\\
    &\quad = h\{\btheta;\mbox{E}_{\bu\vert\bbeta,\btheta;\by}(\bu\bu^\top\,\vert\,\widehat{\bbeta}_{\btheta},\btheta;\by)\}=\bzero.
\end{split}
    \label{eq:profileScoreEquation}
\end{align}
By comparing equations \eqref{eq:scoreEquation} and \eqref{eq:profileScoreEquation}, and using the fact that $\mbox{E}_{\by\vert\bbeta,\btheta}\{\mbox{E}_{\bu\vert\bbeta,\btheta;\by}(\bu\bu^\top\,\vert\,\bbeta,\btheta;\by)\}=\bD(\btheta)$
where the outer expectation is computed with respect to the marginal distribution of $\by$ arising from \eqref{eq:baseGLMMmodel} for known $\bbeta$ and $\btheta$, it follows that the score equation for $\widehat{\btheta}_{\bbeta}$ is unbiased (since $h\{\btheta,\bD(\btheta)\} = \bzero$), while the profile score equation for $\widehat{\btheta}$ is biased downwards due to the estimation of $\widehat{\bbeta}_{\btheta}$. Specifically, we can define
\begin{align*}
    \text{bias}(\bbeta,\btheta)&= \mbox{E}_{\by\vert\bbeta,\btheta}\{\mbox{E}_{\bu\vert\bbeta,\btheta;\by}(\bu\bu^\top\,\vert\,\widehat{\bbeta}_{\btheta},\btheta;\by)\}\\
    &\quad- \mbox{E}_{\by\vert\bbeta,\btheta}\{\mbox{E}_{\bu\vert\bbeta,\btheta;\by}(\bu\bu^\top\,\vert\,\bbeta,\btheta;\by)\},
\end{align*}
and subsequently construct the bias-corrected estimating equation $h\{\btheta;\mbox{E}_{\bu\vert\bbeta,\btheta;\by}(\bu\bu^\top\,\vert\,\widehat{\bbeta}_{\btheta},\btheta;\by) - \text{bias}(\bbeta,\btheta)\}=\bzero$. 
Finally, \citet{liao2002type} suggested substituting the unknown $\bbeta$ with $\widehat{\bbeta}_{\btheta}$, and thus define their REML estimator of $\btheta$ as the solution to the estimating equation
\begin{equation}
    h\{\btheta;\mbox{E}(\bu\bu^\top\,\vert\,\widehat{\bbeta}_{\btheta},\btheta;\by) - \text{bias}(\widehat{\bbeta}_{\btheta},\btheta)\}=\bzero.
    \label{eq:estEquation}
\end{equation}
An estimate of the fixed effects can be then obtained by evaluating $\widehat{\bbeta}_{\btheta}$ at the REML estimate of $\btheta$. Although the estimating equation \eqref{eq:estEquation} is no longer unbiased with the substitution of $\widehat{\bbeta}_{\btheta}$, \citet{liao2002type} argued the dependence of $\text{bias}(\bbeta,\btheta)$ on $\bbeta$ should be weak 
(in fact, they showed the bias does not depend on $\bbeta$ for linear models) 
and thus the difference between $\text{bias}(\bbeta,\btheta)$ and $\text{bias}(\widehat{\bbeta}_{\btheta},\btheta)$ should be small. Practically, \citet{liao2002type} computed the standard maximum likelihood estimators $(\widehat{\bbeta}^\top,\widehat{\btheta}^\top)^\top$ and subsequently the REML estimates via the EM algorithm, while evaluating the bias term in equation \eqref{eq:estEquation} via Monte Carlo methods, e.g.,  averaging $\mbox{E}_{\bu\vert\bbeta,\btheta;\by}(\bu\bu^\top\,\vert\,\widehat{\bbeta}_{\btheta},\btheta;\by) - \mbox{E}_{\bu\vert\bbeta,\btheta;\by}(\bu\bu^\top\,\vert\,\bbeta,\btheta;\by)$ over $m$ clusters in an independent cluster GLMM.

The estimator proposed by \citet{liao2002type} applies to models with normal random effects only, and has the same consistency properties as the maximum likelihood estimator if the number of parameters in the mean and variance components remains fixed. As with the modified profile likelihood class of methods, there are no rigorous theoretical guarantees that using \eqref{eq:estEquation} reduces the leading bias term of the variance component estimates, although the simulation results presented in Section \ref{sec:simStudy} show that the method can effectively reduce the finite sample bias. It is also not hard to see that 
the direct bias correlation method is arguably the most computationally demanding class of REML estimation approaches reviewed so far, since it requires integrating with respect to the conditional distribution of the random effects given the data, estimating the bias via simulation, and finally adjusting for the bias. The computational cost could be reduced by, for example, replacing the Monte Carlo methods with an approximate integration approach.
As a final remark, the method of \cite{liao2002type} is based on a direct bias correction involving adjustments to the estimating equation for GLMMs. There exist other methods which use a bias correction strategy, but without being categorized as a REML approaches \emph{per-se}. For instance, \cite{kuk1995asymptotically} proposed a method for adjusting any conveniently defined initial estimator to obtain an asymptotically unbiased and consistent estimator. This is achieved by adding a simulation-based iterative procedure to correct the bias in the BLUP procedure. 
The proposal of \cite{kuk1995asymptotically} is similar to the algorithm of \cite{liao2002type}, although the former corrects the bias in the estimate itself while the latter correct the bias in the estimating equation.

\section{REML in Hierarchical Generalized Linear Models \label{sec:hglms}}

Hierarchical generalized linear models (HGLMs), introduced by \cite{lee1996hierarchical}, extend the GLMM in \eqref{eq:baseGLMMmodel} by allowing the random effects distributions $p(\bu; \btheta)$ to be non-normal, and in particular to be conjugate with the conditional distribution of the responses given the random effects. Inference in HGLMs is carried out using the hierarchical likelihood, or $h$-likelihood, which is an analogue to the joint log-likelihood function of $\by$ and $\bu$ in \eqref{eq:GLMMjointLogLik} in GLMMs \citep[see][for a recent review of $h$-likelihood and HGLMs]{jin2021review}. For instance, the original $h$-likelihood outlined by \cite{lee1996hierarchical} arises from replacing $\bdeta = \bX\bbeta + \bZ\bu$ in model \eqref{eq:baseGLMMmodel} with a more general linear predictor $\bdeta = \bX\bbeta + \bZ\bv$, where $\bv$ is on an additive scale in the linear predictor, and transformed from the original scale of $\bu$, e.g., using some strictly monotonic function. 
More recently, \cite{lee2023h} proposed a new $h$-likelihood for neural networks with mixed effects, whose joint optimization leads to maximum likelihood estimates for all fixed parameters. Notably, this new $h$-likelihood is not equivalent to the joint density of $\by$ and $\bu$, even for LMMs with Gaussian random effects. Estimation of the variance components in a HGLM is subject to the same bias issues as in GLMMs. Subsequently, a number of REML estimators for the variance components in HGLMs have been proposed, leveraging arguments often analogous to those in Section \ref{sec:approxLinearization} for approximate linearization methods. 

\cite{lee1996hierarchical} developed a REML estimator for HGLMs based on maximization of an adjusted $h$-likelihood, generically denoted here as $h_A = h -(1/2)\log\det(\bM)$, where $\bM$ denotes the expected Hessian matrix from the $h$-likelihood. The REML adjustment $-(1/2)\log\det(\bM)$ is akin to the final terms of \eqref{eqn:wolfinger_REML} and \eqref{eqn:PQL_theta} in the approximate linearization techniques. Moreover, for LMMs it is equivalent to the restricted likelihood of \citet{patterson1971recovery} and the profile log-likelihood correction of \cite{cox1987parameter} for dispersion components when the fixed and random effects are treated as asymptotically orthogonal nuisance parameters. 
For GLMMs with a normally distributed random effects, the adjusted profile $h$-likelihood produces identical REML estimates of the variance components to \cite{schall1991estimation}, \cite{wolfinger1993generalized} and \cite{mcgilchrist1994estimation}. Building on this relationship, \cite{lee2001hierarchical} proposed another REML extension for HGLMs based on a double extended quasi-likelihood, which is applicable to a wider class of models and allows for general random effect covariance matrices, including those of spatial and temporal GLMMs \citep{lee2001modelling}.
More recently, \cite{han2024enhanced} discussed the relationship between the adjusted profile likelihood of \cite{cox1987parameter} mentioned above, the formula for the conditional distribution of the maximum likelihood estimator given the maximal ancillary statistic of \cite{barndorff1983formula} and the restricted likelihood of \cite{lee2001hierarchical}.

In a more detailed examination of REML for HGLMs, \cite{noh2007reml} derived approximate versions of a conditional $h$-likelihood approach to REML. Focusing on the analysis of binary data with crossed random effects, they introduced several extensions of REML from LMMs to HGLMs, and found that these $h$-likelihood methods tend to outperform REML estimators derived using other approaches such as first and second-order Laplace-approximation based methods and the approximate linearization methods of \cite{schall1991estimation}, \cite{breslow1993approximate}, and \cite{lin1996bias}.
Other REML methods identified and criticized by \cite{noh2007reml} for giving too biased estimators of the variance components include those of \citet{drum1993reml}, \citet{shun1995laplace}, who developed a Laplace approximation beyond the first-order (see also \citeauthor{shun1997another}, \citeyear{shun1997another}), and
\citet{lee1996hierarchical,lee2001hierarchical}.
\cite{noh2007reml} also stated that simulation methods such as the Monte Carlo Expectation-Maximization and Monte Carlo Newton-Raphson methods used in developing REML estimators can be computationally intensive and again produce biased estimators.

In another work, \citet{lee2012modifications} studied HGLMs with correlated random effects and show how REML algorithms can be developed using $h$-likelihood methods by modifying the existing ML and REML procedures using adjusted dependent variables for approximate linearization, followed by introducing higher-order Laplace approximations. \cite{yu2013robust} developed another version of the \cite{lee1996hierarchical} estimator using a modified profile likelihood approach which is robust to the impact of extreme random effects, by taking advantage of previous work in \citet{yau2002robust}. Their REML estimator was developed particularly for Poisson GLMMs and then extended to Poisson finite mixture models with random effects; see also the previous work of \cite{wang2002hierarchical}, who used REML estimation for a two-component hierarchical Poisson mixture regression model via the EM algorithm. Furthermore, they confirmed that the original REML estimator of the variance components from \cite{lee1996hierarchical} was particularly biased downwards in finite samples, consistent with the findings of \citet{noh2007reml}.

More recently, \cite{bologa2021high} employed a Laplace-based $h$-likelihood method with a REML correction and implemented via the \textsf{R} package \textsf{glmmTMB} to analyze health records datasets. 
Finally, \cite{han2024enhanced} introduced a REML estimator exploiting an enhanced Laplace approximation which exhibited potential for improved accuracy over the existing methods, but at a higher computational cost, while \cite{jin2024standard} conducted extensive simulation that showed the current standard error estimators are not necessarily accurate, even when the mean parameters are estimated from a higher-order approximation of the likelihood and the dispersion parameters are estimated via REML, and proposed alternative standard error estimators.

\section{Numerical Study \label{sec:simStudy}}

We performed a small simulation study to assess five currently available implementations of REML estimation for independent cluster GLMMs with either binary or count responses, comparing them to standard unrestricted likelihood-based methods in terms of point estimation for variance components. In alignment with our take-home message in Section \ref{sec:Intro}, the choice of the implementations was primarily driven by software availability and ease of implementation in \textsf{R}, but was such that the selected methods broadly represented the four classes of REML estimation methods reviewed in the Sections \ref{sec:approxLinearization} to \ref{sec:biasCorr}, along with REML estimation for HGLMs reviewed in Section \ref{sec:hglms}.
In particular, they include: 
1) approximate linearization via quasi-likelihood estimation, obtained by modifying the function \texttt{glmmPQL} from the \textsf{R} library \textsf{MASS} \citep{venables2002modern} to use equations \eqref{eqn:PQL} and \eqref{eqn:PQL_theta} (by performing the internal call of \texttt{glmmPQL} to the function \texttt{nlme} from the \textsf{R} package \textsf{nlme}, \citeauthor{pinheiro2012nonlinear}, \citeyear{pinheiro2012nonlinear}, via REML, rather than ML as in the original version of the function);  
2) integrated likelihood estimation, via the function \texttt{glmmTMB} from the \textsf{R} package \textsf{glmmTMB} with the REML argument applied, which uses the Laplace approximation as per equation \eqref{eqn:integratedlikelihood_laplace} and noting $\phi$ is fixed to one for binary and count response types; 
3) the modified profile likelihood approach based on equation \eqref{eq:ModProfLogLik}, as implemented in the library \textsf{glmmREML} available at \url{https://ruggerobellio.weebly.com/software.html} at the time of writing, using Gauss-Legendre quadrature to perform the relevant integrations and setting the number of quadrature points in each dimension of the integral to be 50;
4) direct bias correction as defined by the solution to equation \eqref{eq:estEquation}, which was implemented using \textsf{R} code accompanying \cite{liao2002type}, available at \url{http://www.geocities.ws/jg\_liao} at the time of writing, and using 2,000 Monte Carlo samples to perform the relevant integrations; 5) REML for HGLMs performed via the function \texttt{dhglmfit} from the \textsf{R} package \textsf{dhglm} \citep{dhglm2018} with default options. 
Note the implementations of methods 3 and 4 by their respective authors were limited: direct bias correction was only available for independent cluster logistic GLMMs with a random intercept and a random slope, while modified profile likelihood supported this alongside independent cluster Poisson and negative binomial GLMMs with random intercepts only. For this simulation, we also implemented an adaptation of the code of \cite{liao2002type} to perform REML estimation via direct bias correction for independent cluster Poisson GLMMs with random intercepts. 

We compared the five REML estimation approaches for GLMMs to the following unrestricted maximum likelihood estimation methods: using the default `ML' option for \texttt{glmmPQL}, \texttt{glmmTMB} and \texttt{dhglmfit}, and using the available code for unrestricted maximum likelihood estimation provided by \cite{bellio2011restricted} and \cite{liao2002type}. The supplementary material offers further details on the set up for each of the estimation methods outlined above.
In presenting the simulation results, we used the acronyms `PQL', `TMB', `MPL', `DBC' and `HGLM' to refer to the five unrestricted maximum likelihood estimation methods discussed above. Note neither `MPL' or `DBC' actually involve a modified profile likelihood or direct bias correction \emph{per-se} in the case of unrestricted maximum likelihood estimation; rather, we used these acronyms because the code for the corresponding two methods originate from the authors of modified profile likelihood \citep[i.e.,][]{bellio2011restricted} and direct bias correction \citep[i.e.,][]{liao2002type}, respectively. We then paired each of these above acronyms with `REML' to denote the REML estimation defined in methods 1 to 5 above.

\subsection{Binary Responses \label{sec:binSimStudy}}
We generated independent cluster binary data from the logistic GLMM
$y_{ij}\vert\bu_i\sim\mbox{Bernoulli}\big(1/[1+\exp\{-(\bx_{ij}^\top\bbeta+\bz_{ij}^\top\bu_i)\}]\big), \; \bu_i\sim N(\bzero,\bSigma)$, with $i=1,\ldots,50$ clusters and $j=1,\ldots,10$. The above design was employed in both \cite{liao2002type} and \cite{bellio2011restricted} under the following two scenarios, which we subsequently adopted to simulate data:
\begin{itemize}
    \item \underline{Scenario 1}: four fixed effects included as $\bbeta=(\beta_0,\beta_1,\beta_2,\beta_3)^\top=(0.5,1,-1,-0.5)^\top$, with corresponding covariates $\bx_{ij}=(1,x_{ij,1},x_{ij,2},x_{ij,3})^\top$ such that $x_{ij,1}=(j-5)/4$, $x_{ij,2}=0$ for $i=1,\ldots,25$ and $x_{ij,2}=1$ for $i=26,\ldots,50$, and $x_{ij,3}=x_{ij,1}x_{ij,2}$; random effect covariates $\bz_{ij}=(1,x_{ij,1})^\top$; $\bSigma$ is a $2\times 2$ diagonal matrix with $\Sigma_{11}=\Sigma_{22}=0.5$;
    \item \underline{Scenario 2}: this is similar to Scenario 1 except with four additional uninformative fixed effects $\beta_4 = \beta_5 = \beta_6 = \beta_7 = 0$, and corresponding covariates $x_{ij,l}=x_{i,l}$, $l=4,\ldots,7$ drawn independently from the standard normal distribution.
\end{itemize}
We then considered two further simulation designs based on the above:
\begin{itemize}
    \item \underline{Scenario 3}: similar to Scenario 2 except that ten instead of four additional uninformative fixed effects are included, with covariates drawn independently from the standard normal distribution;
    \item \underline{Scenario 4}: similar to Scenario 1 except the random effects covariance matrix has off-diagonal elements $\Sigma_{12} = \Sigma_{21} = 0.25$, i.e., the random intercepts and slopes are correlated.
\end{itemize}
Scenario 3 is expected to be the most challenging one, given a larger number of fixed effects should have a higher impact on the finite sample bias of unrestricted maximum likelihood estimation for the variance components. For each of the four scenarios, we simulated 500 datasets 
and for each dataset we estimated the model parameters using the five selected REML methods and the five unrestricted counterparts. 

Both PQL and PQL-REML generated errors for some of the simulated datasets, and we provide the statistics of this in the supplementary material. There was also one simulated dataset in each of Scenarios 2 and 3 where TMB-REML did not provide estimates of $\bSigma$, along with one simulated dataset in Scenario 4 where HGLMs gave an error. 
The results we report are based only on the available estimates; we believe these convergence issues do not greatly impact the overall conclusions, given results are in line with what anticipated we draw below, and have outlined checks we performed in the supplementary material to verify this.

\begin{figure*}[htb]
\begin{center}
    \includegraphics[width=0.7\textwidth]{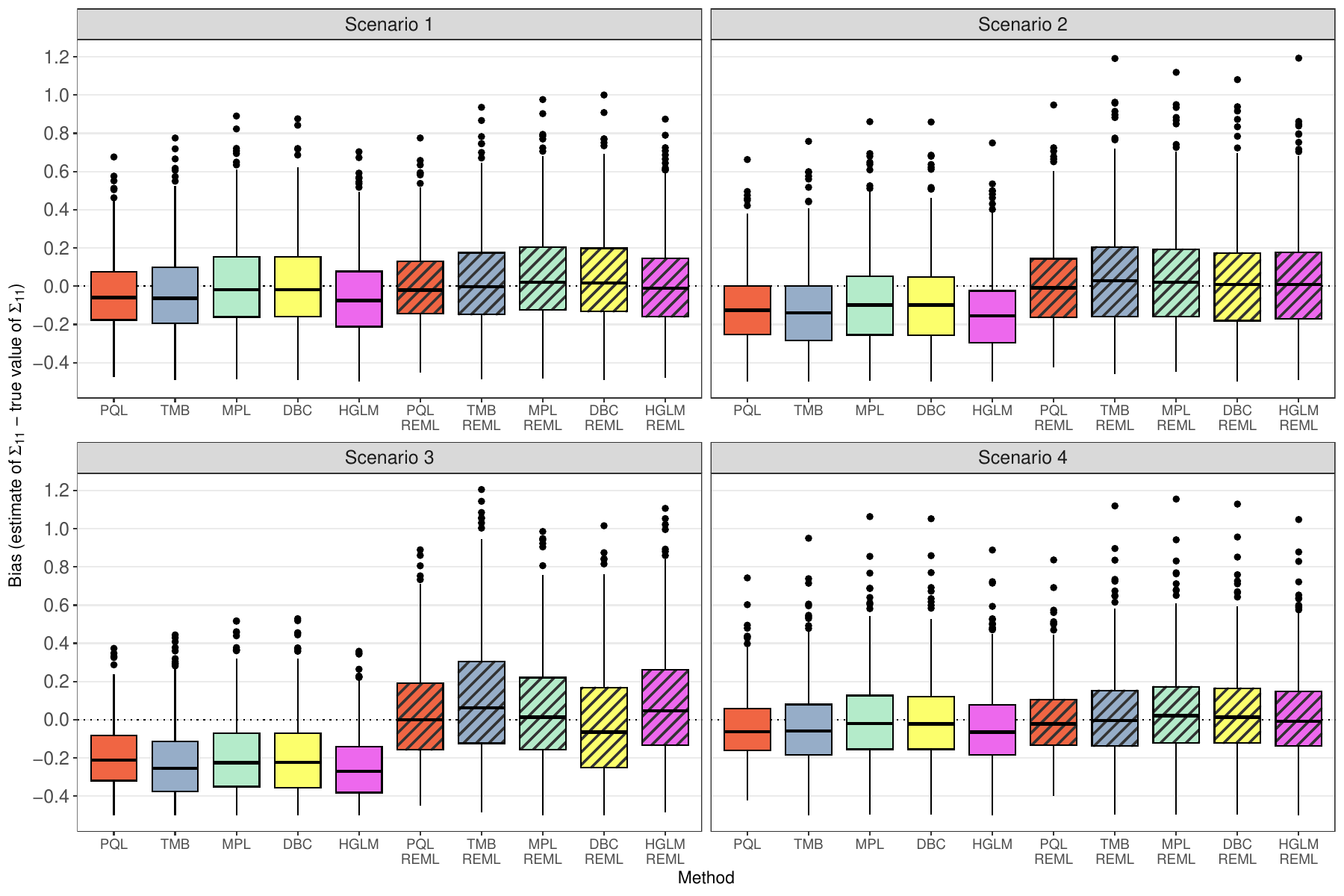}\hfill
    \caption{Boxplots of bias of the estimates of $\Sigma_{11}$ from the simulation study involving binary responses generated from a logistic GLMM. In each panel, the first five boxplots are based on unrestricted maximum likelihood estimation methods, while the remaining five boxplots are based on implementations of REML. Here and in the following figures, the codes 'PQL', 'TMB', 'MPL', DBC' and 'HGLM' respectively refer to estimation performed via the function \texttt{glmmPQL} from the \textsf{R} package \textsf{MASS}, the function \texttt{glmmTMB} from the \textsf{R} package \textsf{glmmTMB}, the code of \cite{bellio2011restricted} for the modified profile likelihood method, the code of \cite{liao2002type} for the direct bias correction method, and the function \texttt{dhglmfit} from the \textsf{R} package \textsf{dhglm}.}
    \label{fig:Sigma11BiasBinWRTtrueSigma}
\end{center}
\end{figure*}
\begin{figure*}[htb]
\begin{center}
    \includegraphics[width=0.7\textwidth]{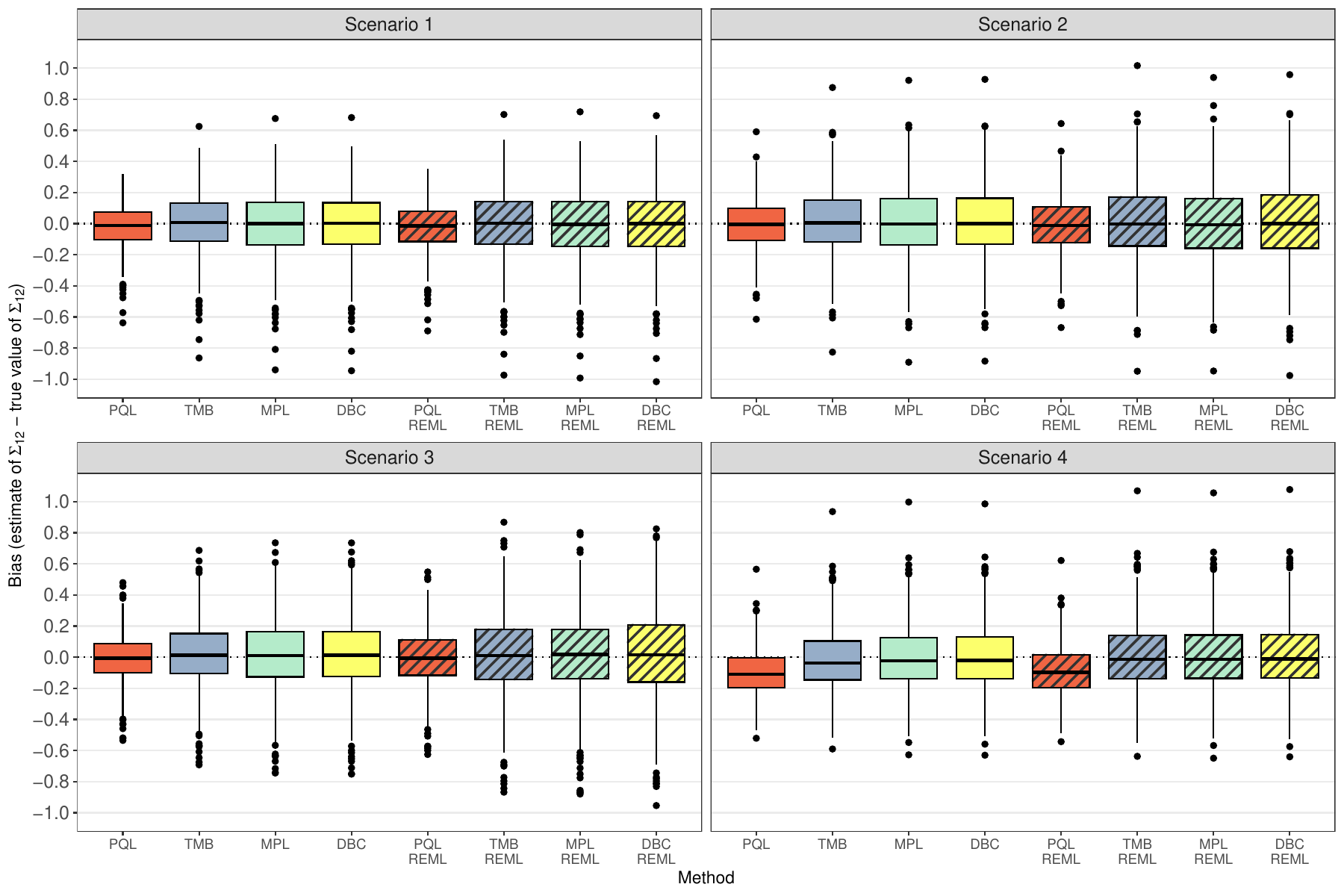}\hfill
    \caption{Boxplots of bias of the estimates of $\Sigma_{12}$ from the simulation study involving binary responses generated from a logistic GLMM. In each panel, the first four boxplots are based on unrestricted maximum likelihood estimation methods, while the remaining four boxplots are based on implementations of REML. The codes have the same meaning as in Figure \ref{fig:Sigma11BiasBinWRTtrueSigma}. Note the function \texttt{dhglmfit} from the \textsf{R} package \textsf{dhglm} does not include an option for correlated random effects, therefore the `HGLM' and `HGLM REML' results are not available for $\Sigma_{12}$.} 
    \label{fig:Sigma12BiasBinWRTtrueSigma}
\end{center}
\end{figure*}
\begin{figure*}[htb]
\begin{center}
    \includegraphics[width=0.7\textwidth]{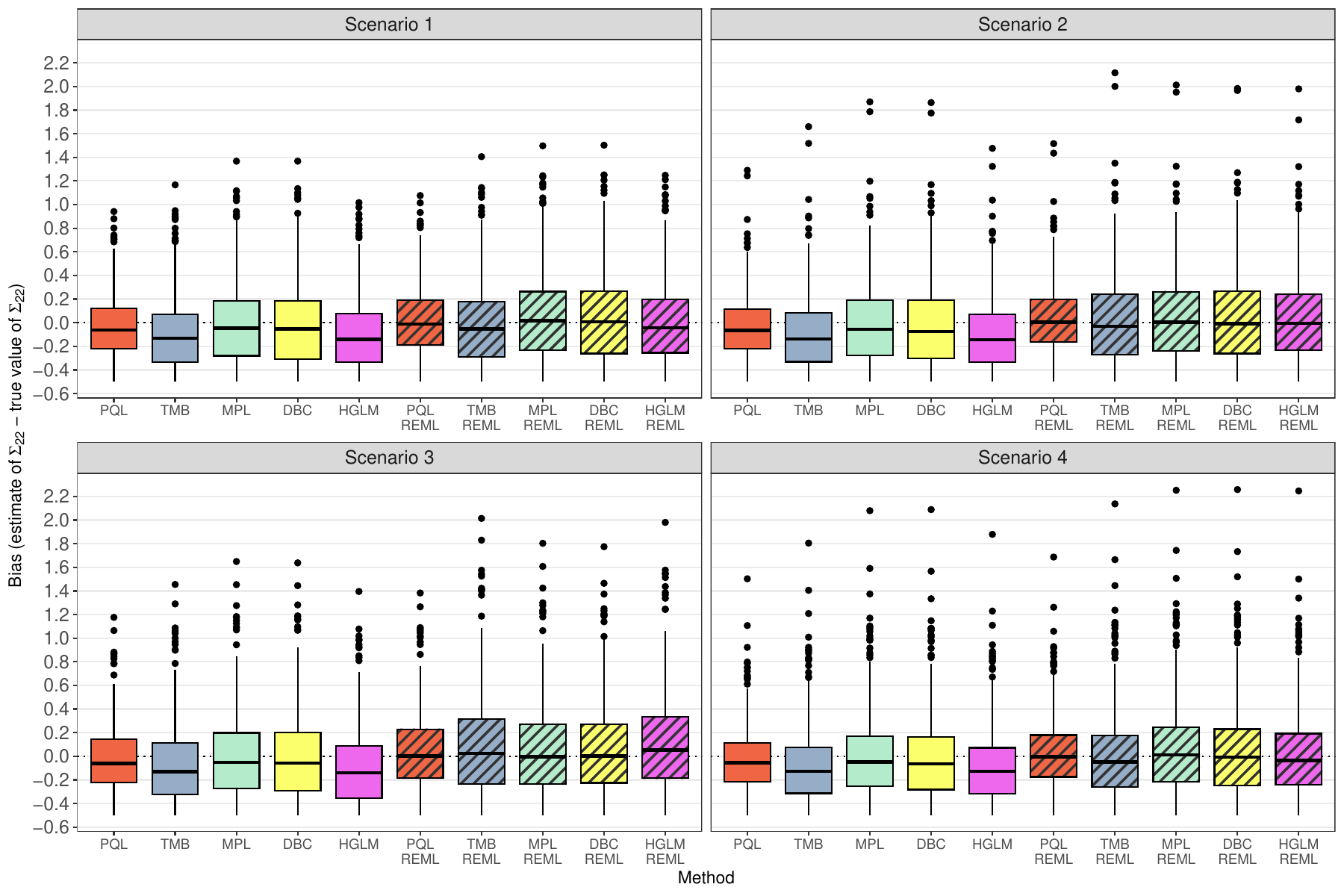}\hfill
    \caption{Boxplots of bias of the estimates of $\Sigma_{22}$ from the simulation study involving binary responses generated from a logistic GLMM. In each panel, the first five boxplots are based on unrestricted maximum likelihood estimation methods, while the remaining five boxplots are based on implementations of REML. The codes have the same meaning as in Figure \ref{fig:Sigma11BiasBinWRTtrueSigma}.
    }\label{fig:Sigma22BiasBinWRTtrueSigma}
\end{center}
\end{figure*}

Figures \ref{fig:Sigma11BiasBinWRTtrueSigma}--\ref{fig:Sigma22BiasBinWRTtrueSigma} present boxplots of the differences between the estimates of $\Sigma_{11}$, $\Sigma_{12}$ and $\Sigma_{22}$ and the corresponding true values, noting that the current implementations of HGLM and HGLM-REML in \texttt{dhglmfit} do not estimate $\Sigma_{12}$, and so the function was applied assuming the random intercepts and slopes were independent.  
All five REML methods proved to be effective in reducing the finite sample bias of the estimates for $\Sigma_{11}$ and $\Sigma_{22}$, when compared with their unrestricted counterparts. This is particularly evident in the boxplots of the random intercept variance $\Sigma_{11}$ and Scenarios 2 and 3. However, the reduction in bias is traded off by the REML estimates tending to generally exhibit more variability compared to their unrestricted alternatives. Comparing the four unrestricted likelihood-based methods only, MPL and DBC tended to produce slightly smaller average biases 
than PQL, TMB and HGLM. A possible explanation for this may be MPL and DBC utilizing numerical/more complex quadrature methods, 
compared to the functions \texttt{glmmmPQL}, \texttt{glmmTMB} and \texttt{dhglmfit} which use faster but simpler analytical approximation strategies for overcoming the integral in the marginal log-likelihood of the GLMM. For $\Sigma_{22}$ in particular, the TMB and HGLM estimates exhibited consistently more bias than other unrestricted likelihood methods across all four scenarios.

Of the five REML methods, MPL-REML performed consistently well for $\Sigma_{11}$ and $\Sigma_{22}$ across all scenarios, while DBC-REML showed potential under correction for the bias of $\Sigma_{11}$ in Scenario 3. Both PQL and PQL-REML generated smaller ranges of bias values, although these results must be interpreted cautiously due to the aforementioned convergence issues with these methods, and noting that the variance components in this simulation are not specially large (meaning the known issues around parameter estimation bias for PQL, as mentioned in Section \ref{subsec:pql}, are less relevant here).

Finally, for the off-diagonal parameter $\Sigma_{12}$, all unrestricted maximum likelihood methods, excluding PQL in Scenario 4 and HGLM which did not provide results for $\Sigma_{12}$, were prone to less bias; this is not surprising given in Scenarios 1 to 3 the true value of $\Sigma_{12}$ is zero. A similar pattern was observed for the four REML methods available for $\Sigma_{12}$, with PQL-REML being less effective than the other REML methods on average.

In the supplementary material, we present results for the differences between each estimate and `exact' profile likelihood estimates of $\bSigma$ computed at the true $\bbeta$ value using the same Monte Carlo approach of DBC and DBC-REML. \cite{liao2002type} and \cite{bellio2011restricted} referred to these as the `ideal' estimates, on the grounds that they are the best possible estimators of the variance components one can obtain assuming the fixed effects are known.
Results from these exhibit similar trends as in Figures \ref{fig:Sigma11BiasBinWRTtrueSigma}--\ref{fig:Sigma22BiasBinWRTtrueSigma}, only  with different boxplot spreads. 
The supplementary material also provides simulation results for the root mean squared error, results for the estimation of the fixed effects $\bbeta$, and a summary of computational times. Consistent with findings in the literature \cite[e.g.,][]{lee2017generalized}, PQL tended to produce more biased estimates of $\bbeta$ than other approaches, in particular HGLMs. Meanwhile, the computational times for MPL-REML and DBC-REML were on the order of minutes to run on a personal laptop, as opposed to fractions of seconds for PQL-REML and TMB-REML and a few seconds for HGLM-REML, keeping in mind MPL-REML and DBC-REML were implemented using code accompanying journal articles, while the last three were run using established \textsf{R} packages.

\subsection{Count Responses}

We simulated independent cluster count data from a Poisson GLMM model with a random intercept and the canonical log link $y_{ij}\sim\mbox{Poisson}\big\{\exp(\bx_{ij}^\top\bbeta+u_i)\big\},\; u_i\sim N(0,\sigma^2)
$, where again we considered $i=1,\ldots,50$ clusters and $j=1,\ldots,10$. To reiterate, we considered only a random intercept in this simulation setting since MPL and MPL-REML have code publicly available only for the random intercept case for Poisson GLMMs. With this in mind, we used Scenarios 1--3 as described in Section \ref{sec:binSimStudy} but with the random effects covariance matrix $\bSigma$ replaced by $\sigma^2=0.25$, and defined a new Scenario 4 using the setting of Scenario 1 but with $\beta_0$ set to $-0.5$ instead of $0.5$ to produce a more challenging low-count data setting. We generated 500 datasets for each scenario, and compared the same eight estimation methods discussed in the preceding subsection. Both MPL-REML and DBC-REML generated errors for some of the simulated datasets (see the supplementary material for details), and below we report results based only on the available estimates.

\begin{figure*}[tb]
\centering
    \includegraphics[width=0.7\textwidth]{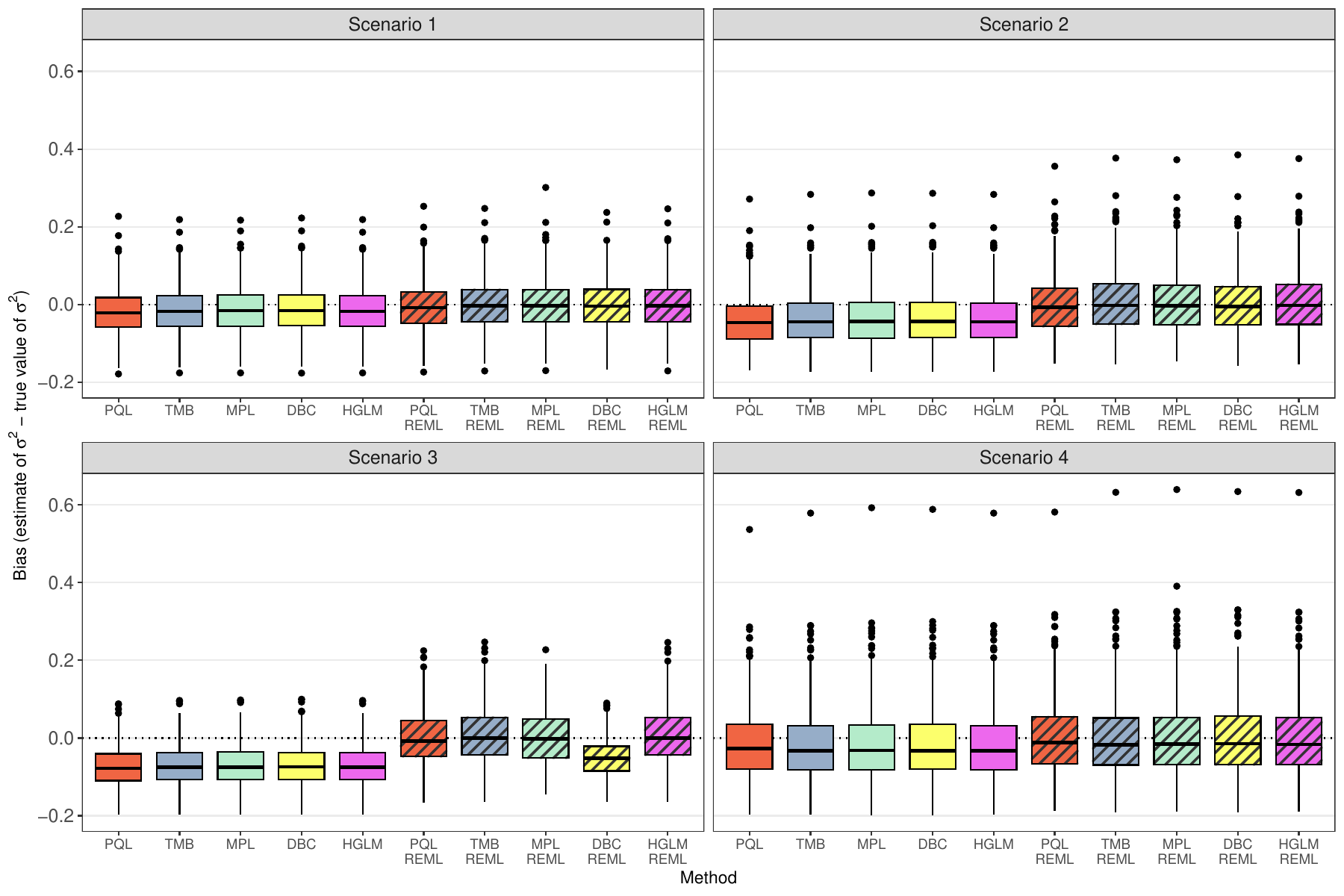}\hfill
    \caption{ Boxplots of bias of the estimates of $\sigma^2$ from the simulation study involving count responses generated from a Poisson GLMM. In each panel, the first five boxplots are based on unrestricted maximum likelihood estimation methods, while the remaining five boxplots are based on implementations of REML. The codes have the same meaning as in Figure \ref{fig:Sigma11BiasBinWRTtrueSigma}.}
    \label{fig:sigsqBiasPoisWRTtruesigsq}
\end{figure*}

Figure \ref{fig:sigsqBiasPoisWRTtruesigsq} presents boxplots of the differences between the estimate and true value of $\sigma^2$. All five unrestricted maximum likelihood methods performed consistently within each scenario, with bias increasing as the number of fixed effect covariates increased, i.e., moving from Scenarios 1 to 3. Consistent with the binary response case results for $\Sigma_{11}$, all REML methods provided a notable bias correction with the exception of DBC-REML in Scenario 3. Regarding the latter, its poor performance is likely related to the aforementioned instability issues for a non-negligible proportion of the simulated datasets. The bias adjustment of PQL-REML was slightly less pronounced than those of TMB-REML, MPL-REML and HGLM-REML. Finally, compared with Scenario 1, the low-count Scenario 4 gave wider ranges of bias values for all unrestricted and REML methods, and the average biases of the REML methods were less close to the zero target. This is not too surprising and is consistent with the idea that low-count responses pose a greater estimation challenge for all REML estimation methods considered here. 

Boxplots of the differences between each estimate and the `exact' estimate based on the profile likelihood of $\sigma^2$ assuming $\bbeta$ is known, along with results for root mean squared error, for the fixed effects $\bbeta$ and summaries of computational times, are provided in the supplementary material. These additional results exhibit largely similar trends to those in seen for binary response in the preceding section, e.g., PQL tended to produce the most biased estimates of the fixed effect coefficients, and MPL-REML and DBC-REML were by far the slowest of the methods examined.

Finally, in additional simulation studies (not shown) for both binary and count responses, we assessed estimation performance with a smaller ($m = 25$) and larger ($m = 100$) number of clusters, while keeping cluster size fixed to 10. As expected, compared to the results presented above, the biases of all unrestricted maximum likelihood methods were larger (smaller) when $m = 25$ ($m = 100$). Applying any of the REML methods tended to reduce these biases, although at $m = 25$ all REML methods tended to exhibit a non-negligible amount of bias.

\section{Discussion \label{sec:lastSection}}
This article highlights the different paths that arise from attempting to translate the concept of restricted maximum likelihood or REML estimation into GLMMs.
We defined four main classes of REML methods in the current GLMM literature -- approximate linearization, integrated likelihood, modified profile likelihood and direct bias adjustment for the score function, and also reviewed methods for REML estimation in the related but broader framework of hierarchical generalized linear models. As hinted throughout Sections \ref{sec:approxLinearization} to \ref{sec:biasCorr} however, our proposed categorization is not univocal, and a central finding of this review is recognizing previously under-stated similarities, and sometimes equivalences, between the different categories of REML estimation, with differences arising mainly through motivation and implementation. 
For instance, the class of approximate linearization methods including \cite{schall1991estimation} and \cite{breslow1993approximate} involve some form of adjustment to a profile likelihood function, whereas 
the proposal of \cite{bellio2011restricted} explicitly adjusts the GLMM profile likelihood function without introducing an initial linearization step. 
Ties between the approximate linearization approach of \cite{schall1991estimation} and the integrated likelihood of \cite{stiratelli1984random} were outlined in the former work itself, although this seems to have been largely been forgotten as these two classes of REML estimation methods subsequently forged their own paths in the literature; see also \cite{wolfinger1993laplace} in the context of nonlinear models.
Finally, both the methods of \cite{bellio2011restricted} and \cite{liao2002type} aim at reducing bias in the score function, with the former through a modification of the profile likelihood function motivated by asymptotic arguments,
while the latter via subtraction of an explicit expression of bias to the score function.  

In line with the above, our simulation study suggests currently available implementations of REML estimation in \textsf{R} perform quite similarly at reducing bias of the variance component estimates, compared with their unrestricted alternatives.
Overall then, given these promising numerical results, from an applied standpoint we encourage REML estimation be more widely employed in the GLMM context if the aim is to reduce bias in estimation of variance components. Moreover, as stated in Section \ref{sec:Intro}, one of our key messages is that the choice of which REML implementation to use should, at least at the time of writing this review, be driven by practical factors such as ease of implementation for a specific GLMM, and computational scalability.
For instance, both the REML versions of \texttt{glmmPQL} and \texttt{glmmTMB} allow for application of REML estimation to a variety of GLMM formulations, although the former was prone to convergence errors while current documentation for the latter advises the practitioner to carefully assess the suitability of the REML option prior to real application. By contrast, the current implementations of the modified profile likelihood and direct bias correction methods are computationally intensive and/or are not straightforwardly generalizable beyond the independent cluster setting. 

As an aside, it is worth acknowledging that in our simulation studies the REML bias correction for GLMMs came at the cost of increased sampling variability compared to standard unrestricted maximum likelihood estimation. While not too surprising, this tradeoff between bias and variance seems to largely be overlooked when choosing between REML or unrestricted estimation methods. We encourage the drawback of increased sampling variability be kept into account by practitioners in their application of GLMMs.

One clear and consistent gap we encountered throughout the review is the general lack of theoretical justification for the approaches to REML estimation for GLMMs. This is not too surprising given the majority of REML methods originate from extension of one of many alternative derivations of REML in LMMs, and in the linear case all derivations result in the same variance component estimates, while the identity link and normal error terms readily facilitates asymptotic if not exact investigation of their properties \citep{richardson1994asymptotic}.
By contrast, we conjecture that the non-identity link and non-constant mean-variance relationship in GLMMs means any proposed REML estimation method at best removes the leading bias term. More broadly, while there exist some theoretical arguments supporting the notion of REML estimation resulting in less biased estimates of the variance components (see Section \ref{subsec:pql} for the large sample justification of PQL due to \citeauthor{wood2017generalized}, \citeyear{wood2017generalized}, Section 3.4.2), these are currently piecemeal and not rigorous. Therefore, a key conclusion of this review is that more formal theoretical research into REML estimation for GLMMs is urgently needed; for example, under what regularity conditions can REML remove the leading bias term, or can we compare finite sample probability bounds on the error norm for REML estimates of the variance components versus its unrestricted counterpart? Perhaps such investigation may be able to make use of
the growing series of theoretical works involving likelihood-based GLMM estimation. At their respective time of development, works such as those of \cite{schall1991estimation}, \cite{stiratelli1984random} and \cite{liao2002type}, among others, lacked existing asymptotic results which they could leverage to theoretically study their proposed REML methods. Over the past decade, however, there has been a growing series of works focusing on more precise large sample results for likelihood-based GLMM estimation 
\citep{jiang2013subset,jiang2022usable,maestrini2024,ning2024asymptotic}. As such, we believe a more formal theoretical examination of REML estimation for GLMMs, and the consequence for inference, is now increasingly possible and we encourage it as an avenue for future research. 

The primary focus of this article has been on bias correction for variance components in GLMMs using methods which have referred to themselves as a form of REML estimation. Alternative ``non-REML" techniques however have also been proposed to solve the bias issue, and a formal comparison of REML methods versus these alternative strategies for bias correction warrants further investigation. For example, \cite{jiang1998consistent} proposed estimators of GLMM fixed effects and variance components using simulated moments, and showed that the biases of these estimators were substantially smaller than those of the approximate REML estimators of \cite{mcgilchrist1994estimation}, although this came at the cost of larger standard errors.
Other general bias reduction strategies for GLMMs could be inspired, for instance, by works such as 
\cite{kosmidis2009bias}, where the main idea is to obtain suitable modifications of score functions whose resulting estimators have smaller bias than the original estimators.

%
%

\begin{acks}[Acknowledgments]
 The authors express gratitude to the three anonymous referees for their helpful comments that improved the quality of the manuscript.
\end{acks}
\begin{funding}
This work was supported by the Australian Research Council Discovery Early Career Research Award\\ DE200100435 and the Australian Research Council Discovery Project DP230101908.
\end{funding}

%
\begin{supplement}
\stitle{Marginal Quasi-Likelihood Estimation for GLMMs}
\sdescription{Description of marginal quasi-likelihood estimation and its REML adaptation in GLMMs.}
\end{supplement}
\begin{supplement}
\stitle{Other Approximate Linearization Methods for REML
Estimation in GLMMs}
\sdescription{Review of other approximate linearization methods for REML estimation in GLMMs.}
\end{supplement}
\begin{supplement}
\stitle{Additional Details for the Simulation Study}
\sdescription{Additional simulation study details, estimation results about variance components and fixed effects parameters, and summaries of computational time.}
\end{supplement}
%


\bibliographystyle{imsart-nameyear}
\bibliography{ref}       


\includepdf[pages={1-19}]{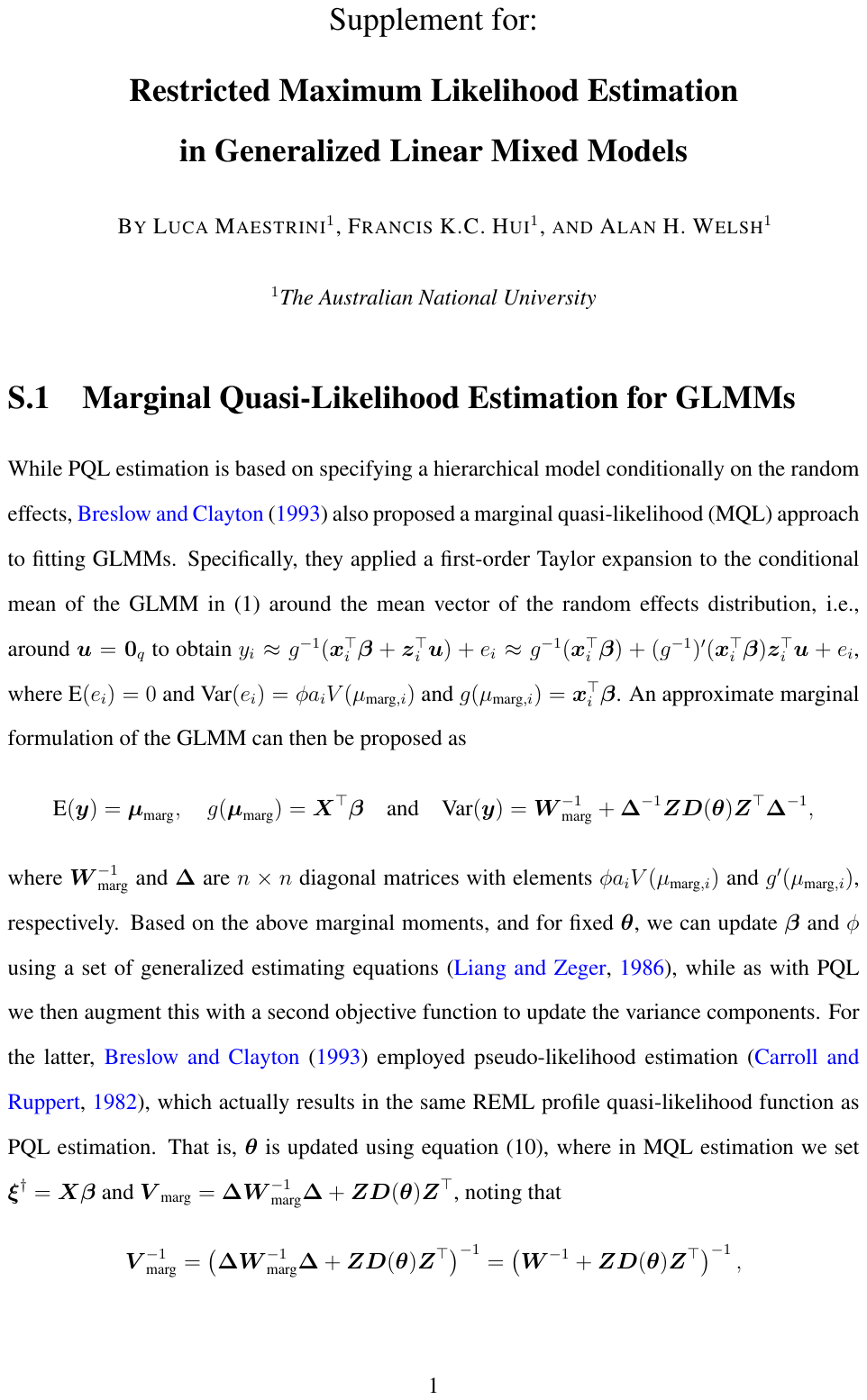}

\end{document}